\documentclass[manuscript]{aastex}
\slugcomment{}

\shorttitle{IR compilation of Bulge GCs}
\shortauthors{Valenti et al.}

\begin{document}

\title{Near--IR properties of 24 Globular Clusters in the Galactic bulge
\footnote{Based on data taken at the ESO/NTT Telescope, within the observing programs:
73.D--0313, 75.D--0372 and 77.D--0757.}}

\author{E. Valenti}
\affil{
ESO -- European Southern Observatory, Alonso de Cordoba~3107, Casilla
19001, santiago~19, Chile\\
INAF -- Osservatorio Astronomico di Bologna, Via Ranzani~1, I--40127
Bologna, Italy\\}
\email{evalenti@eso.org}
\author{F.~R. Ferraro} 
\affil{Bologna University, Via Ranzani~1, I--40127
Bologna, Italy}
\and
\author{L. Origlia}
\affil{INAF -- Osservatorio Astronomico di Bologna, Via Ranzani~1, I--40127
Bologna, Italy}

\begin{abstract}
We present  near--IR   
Color--Magnitude Diagrams and 
physical parameters for a sample
of 24 Galactic Globular Clusters toward the Bulge direction. In this  paper 
we discuss the properties of 
twelve new clusters (out of 24) in addition to those 
previously studied and published by our group.
The compilation includes measurements of the cluster reddening,
distance, photometric metallicity, Horizontal Branch Red Clump, 
Red Giant Branch morphological (i.e.
mean ridge lines) and evolutionary (i.e. bump and tip) features.
The compilation is available in electronic form through the
WorldWideWeb, and it will be updated regularly.
\end{abstract}

\keywords{globular clusters: general --- Galaxy: bulge --- infrared: stars ---
stars: imaging --- technique: photometric}

\section{Introduction}
In the last decades
Galactic Globular Clusters (GCs) have proven to be extremely
important astrophysical laboratories for a wide range of problematics. 
Indeed, the study of their stellar populations address
fundamental questions ranging from stellar structure, evolution and
dynamics to Galaxy formation and the early epoch of the
Universe. Containing some of the oldest stars known, they are {\it fossils} from the 
remote and violent epoch of Galaxy formation. They also serve as
test particles for studying Galaxy dynamics and to test stellar
dynamical models. Being the largest aggregates in which all post Main Sequence (MS) 
stars can be individually observed, they serve as fiducial templates for
understanding the integrated light from distant, unresolved stellar systems. 
In this respect, the
Bulge GCs provide ideal templates to explore the high metallicity regime and thus to
study the stellar content of extra--galactic bulges and Ellipticals. 
However, the high and patchy extinction, which makes optical observations difficult
to impossible along most Bulge lines of sight, coupled with the limited performances 
of the
past generation of near--IR instrumentation, prevented
accurate determinations of the Bulge GCs basic properties 
(as age, metallicity, distance,
etc), as already existing since a long time in the case of the Halo clusters.
The current generation of ground--based IR instrumentation
with high spatial resolution and wide field coverage, and with the
future availability of the James Webb Space Telescope will allow us to resolve the
brightest giants in galaxies up to several Mpc away. Hence, 
a homogeneous compilation of the Bulge GCs properties to be used as empirical
templates of metal--rich stellar populations, is thus strongly urgent. 
In this framework, our group started a long--term project devoted to fully characterize
the stellar populations in the Bulge GC system, by using color--magnitude diagrams
(CMDs) and luminosity functions (LFs) in the near--IR (see
\citet{frf00} -- hereafter F00; \citet{v04}; \citet{vfo04a} -- hereafter VFO04a;
\citet{vfo04b} -- hereafter VFO04b; \citet{vof05} -- hereafter VOF05; 
\citet{ori05} -- hereafter O05). 
The collected photometric database
has been used to perform a detailed description of the main morphological and
evolutionary features of the Red--Giant Branch (RGB) sequence, by means of a set of
photometric indices, which have been defined and widely described in F00, and
calibrated as a function of the cluster metallicity in VFO04a and VFO04b. 

In this paper we present the largest, homogeneous near--IR photometric database of Bulge 
GC ever obtained. For each clusters, the compilation includes: 
{\it i)} the photometric catalog; 
{\it ii)} the RGB mean ridge line,; {\it iii)} accurate reddening, distance, and metallicity
determinations; {\it iv)} the luminosity of the
main RGB evolutionary features (i.e. the bump and the tip); and {\it v)}
the mean magnitude of the Horizontal--Branch Red Clump (HB--RC). Moreover, 
in order to be easily accessible by the community, the compilation would be
also provided in electronic form.

The programme cluster sample is presented in \S~2, while \S~3 is devoted to the
detailed description of the observations, data reduction, and derived CMDs of a
sample of 12 Bulge clusters, recently observed, and presented here for the first time.
\S~4 describes the overall characteristics of the compilation and the measurement
of the HB--RC and the RGB bump and tip.

\section{The entire sample}
\begin{figure}
\plotone{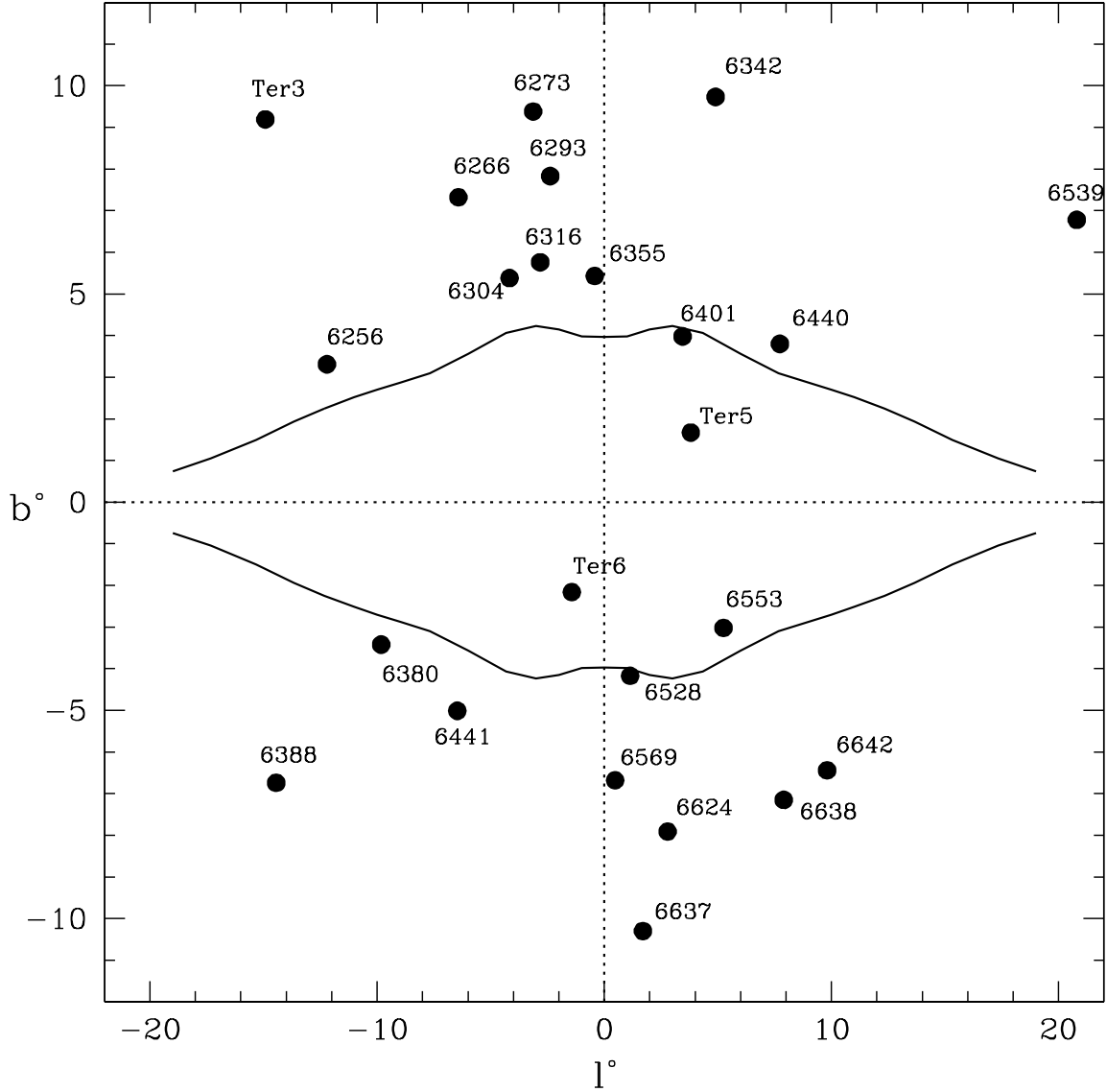}
\caption{\label{map} Position of the Bulge clusters global sample with 
respect to the COBE/DIRBE
3.5$\mu$m inner Bulge outline ({\it solid line}, \citet{cobe}) at 5 MJy
sr$^{-1}$.}
\end{figure}

The cluster sample presented in this study counts 24 GCs in the Bulge
direction observed in the near--IR by our group in the last few years.
The targets selection has been performed by defining 
{\it Bulge GCs} all those located within $\rm |b|{\leq}10^{\circ}$ and 
$\rm |l|{\leq}20^{\circ}$, where {\it l} and {\it b} are the
Galactic coordinates, and giving a highest priority to the metal--rich population.
Note that this definition is mainly a working hypothesis
and refers to the position of the clusters in the Bulge direction, only. However, there
are a number of growing evidences, based on kinematics \citep{din03} and
high--resolution chemical abundances \citep[][ and references therein]{origlia05},
favoring a Bulge origin for the metal--rich GCs within 3~Kpc from the Galactic center.
Fig.~\ref{map} shows the spatial distribution in Galactic coordinates of our
sample (which counts 50\% of the entire Bulge GC system),
and the COBE/DIRBE 3.5$\mu$m inner Bulge outline \citep{cobe}. 
The clusters discussed in this paper fall into three categories:

{\bf Sample A} -- 
{\underline {Seven clusters}} (NGC6342, NGC6380, NGC6440, NGC6441, NGC6528, NGC6553, 
NGC6624) have been already published, and 
a detailed description of the observations, data reduction and analysis can be found in
F00, VFO04a, and VFO04b.
This sample have been used to perform a detailed analysis of the main RGB
morphological and evolutionary features, leading to
an empirical calibration of suitable near--IR photometric indices (i.e. RGB color,
magnitude, slope, bump and tip) as a function of the cluster metallicity.

{\bf Sample B} -- {\underline {Five clusters}} (NGC6304, NGC6569, NGC6637,
NGC6638, and NGC6539) have been presented in VOF05 and O05,
where a detailed description of the derived IR--CMDs and the analysis of the RGB
properties can be found. 
For this sample we derived the clusters metallicity by using
the calibration of VFO04a,b.

{\bf Sample C} -- {\underline {Twelve clusters}} (NGC6256, NGC6266, NGC6273, NGC6293,
NGC6316, NGC6255, NGC6388, NGC6401, NGC6642, Ter3, Ter5, Ter6) are  
presented and discussed here for the first time. 

\section{Properties of Sample C}
For this sample, 
a detailed description of the observations, data reduction, CMDs and physical properties of each
cluster is provided in the following  sub--sections.
 
\subsection{Observations and data reduction}
J, H and K$_s$ images of the clusters in {\it Sample C}  
were obtained during three observing runs
on June 2004, July 2005, and June 2006 using the near--IR camera
SofI, mounted at the ESO/NTT telescope. 
During the observing runs two set of data were secured:

{\it (i) Standard resolution set.} A series of images in the J, H and K$_s$ bands
have been obtained by using SofI in {\it Large Field mode},
characterized by a pixel size of 0.288" and a total field of view of 
4.9'$\times$ 4.9'. On average, the images are the combination of 42, 72, and 99
exposures each one 3--sec long in the J, H and K$_s$ passbands,
respectively.

{\it (ii) High resolution set.} High resolution images of the inner region of
each cluster were also secured using
the SofI focal elongator, yielding a pixel size of 0.146" and a total field of
view 2.49' $\times$ 2.49'. High resolution images are the average of
30 single exposures 1.2 sec--long.
All the secured images are roughly centered on the cluster center.

Note that the region covered by our observations allows us to sample 
a significant fraction of the total cluster light
(typically ${\sim}$80-95\%) in all the programme clusters. 
During the 8 nights of observations 
the average seeing was always quite good (FWHM${\approx}0.8"-1"$).
Every image has been background--subtracted 
by using sky fields located several arcmin away from the cluster center, 
and flat--field corrected 
using halogen lamp exposures, 
acquired with the standard SofI calibration setup. 

Standard crowded field photometry, including PSF modeling, was carried out 
on each 
frame by using DAOPHOTII/ALLSTAR \citep{dao94}. 
For each cluster, two photometric 
  catalogs (derived from high and standard resolution images), 
  listing the instrumental 
  J, H and K$_s$ magnitudes, were obtained by cross-correlating the 
  single-band catalogs. 
  The standard and high resolution catalogs have been
  combined by means of a proper weighted average, weighting more the 
  high resolution measurements in the innermost region of the cluster.
  In principle, this strategy allows to minimize the blending effects. 
  The internal photometric accuracy 
  has been estimated from the {\it rms} frame--to--frame scatter of multiple
  stars measurements. Over most of the RGB extension, the internal errors are
  quite low (${\sigma}_J{\sim}{\sigma}_H{\sim}{\sigma}_K<0.03$ mag), increasing
  up to ${\sim}0.06$ mag at $K_s{\geq}16$.

The instrumental magnitudes were then converted 
into the 2MASS photometric system{\footnote {
An overall uncertainty of ${\pm}0.05$ mag in the zero--point 
calibration in all three bands has been estimated.}}, and the
star positions astrometrized onto 2MASS{\footnote {
The astrometric procedure provided {\it rms} residuals of 
$\approx$0.2 arcsec in both R.A. and DEC.}}
as done for the clusters in {\it Samples A and B}. 

Since for Terzan~6 only high resolution data were acquired, 
we also used
J, H and K' images obtained with the IRAC--2 camera (mounted at the 
ESO/MPI~2.2m telescope), in order to cover a larger area ($\approx$ 4'$\times$4'). 
The derived calibrated catalog has been then
merged with that one obtained from the SofI observations.
Note that, as already  discussed in \citet{v04} the calibration of the
K' IRAC--2 photometry onto the K$_s$ SofI one requires negligible color term.

Figure ~\ref{cmds} presents the derived IR CMDs of the global sample of 24 GCs. As shown
in the figure, the photometric catalogs span the entire RGB extension, from the Tip to
$\approx$2\--5~magnitudes below the HB (depending on the cluster extinction).

\subsection{Reddening, distance and metallicity}
The main cluster properties such as reddening and distance have
been derived by using a differential method based on the comparison of CMDs 
and LFs of clusters with similar HB
morphology. However, this method allows to derive reliable estimates of reddening and
distance only for intermediate--high metallicity clusters, whose CMD shows a red clumpy HB morphology. In
fact, the location in magnitude of the HB--LF peak can be used as a reference feature 
and safely compared with
that one of the template cluster.
In 5 metal--poor clusters  
(i.e. NGC6256, NGC6273, NGC6293, NGC6355 and NGC6401) the blue HB morphology coupled with 
the relatively high reddening
and the remarkable level of field contamination prevented a safe location of the HB level,
and thus the use of the differential method. Hence, in these cases the cluster
reddening and distance have been obtained by using the empirical method presented
by \citet[][-- hereafter FVO06]{fvo06}, which allows one to simultaneously get reddening, distance and
metallicity of a stellar system by using a few observables, like the RGB slope, tip,
and mean ridge line, of the [K, J--K] CMD. 

The derived cluster reddening and distance have been used to transform 
the observed CMDs and RGB ridge lines into the
absolute plane, and to measure the following parameters:
{\it i)} (J--K)$\rm _0$ and (J--H)$\rm _0$
colors at 4 fixed absolute magnitude levels ($\rm M_K=M_H=-5.5,-5,-4,-3$); 
{\it ii)} the absolute $\rm M_K$ and $\rm M_H$ magnitudes 
at constant (J--K)$\rm _0=$(J--H)$\rm _0=0.7$ colors; and {\it iii)} the slope in the [K,
J--K] and [H, J--H] planes.
Then, by using the empirical calibrations from VFO04a linking this set of
photometric indices to the cluster metal content, we finally derived the
photometric metallicity estimates in the \citet{cg97} scale. Hence, hereafter the
notation [Fe/H] refers to the \citet{cg97} scale. 
Note that, to derive the global metallicity [M/H], which takes into account the iron 
as well as the $\alpha$--element
abundances, we used the calibrations presented in FVO06 within the
Bulge\--like enrichment scenario. Hence, accordingly to Carney's (1996) suggestions and
the more recent results on the bulge populations 
(\citet{origlia02,origlia04,zoc04,origlia05,ori05,eug06} for bulge clusters;
and \citet{ro05,cun06,manu06,fmr06} for giant field stars) we
adopted a [$\alpha$/Fe]=0.30 constant over the entire range of metallicity up to solar
[Fe/H].

In the next paragraphs we present the CMDs and RGB fiducial ridge lines for the 12 Bulge
clusters of {\it Sample~C}. We briefly discuss the main CMD properties, and we provide
some references for previous published photometries. 

\subsection*{NGC~6256}
NGC~6256 is a heavily reddened cluster in the outer Bulge region. 
Its extinction, whose estimates range from E(B--V)=0.84 \citep[][hereafter H96]{harris96} to
E(B--V)=1.66 \citep[][hereafter S98]{schlegel98}, prevented detailed photometric studies in the
optical. Only two studies of the cluster can be found in the
literature. The first by \citet{barb98} based on VI ground--based photometry,
and the second by \citet{ptt02} based on HST BV observations. 
However, the derived CMDs show an extremely scattered RGB which is
barely defined.

\begin{figure}
\epsscale{1.1}
\plotone{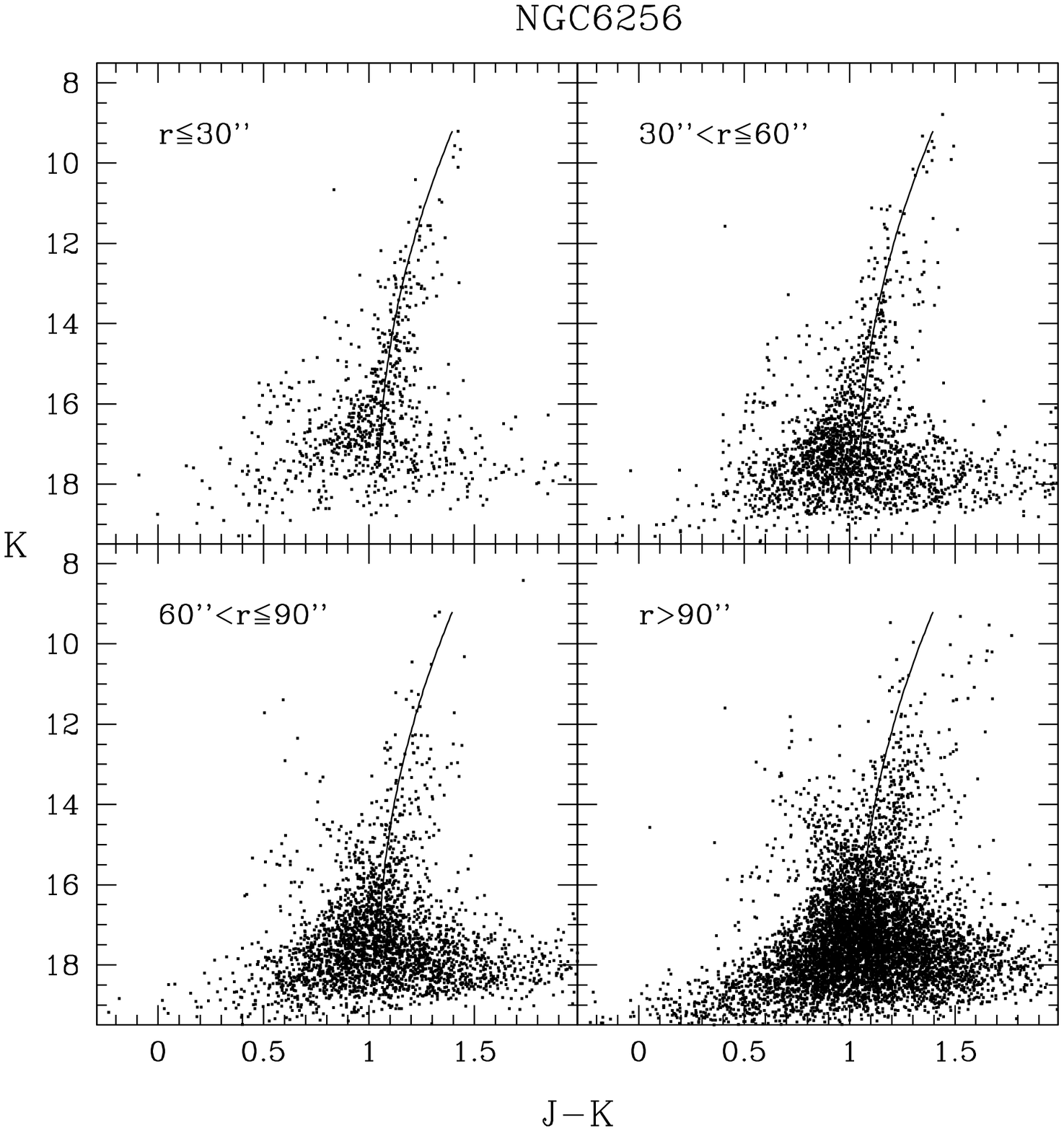}
\caption{\label{6256rad} [K, J--K] CMDs of NGC~6256 at
different distances (r) from the cluster center.}
\end{figure}

In our IR\--CMD, shown in Fig.~\ref{cmds}, 
the cluster RGB
is rather scattered, suggesting the presence of a red component which might be
due to the contamination by Bulge field stars. The HB appears
as a blue vertical structure at (J--K)$\sim$0.5 and K$\leq$14 typical of
low--intermediate metallicity. The almost blue vertical
sequence, at 11.5$\leq$K$\leq$15 and 0.4$\leq$(J--K)$\leq$1, is likely
due to the foreground disk stars. 
In order to investigate the nature of
the observed RGB spread, we compared the 
sequence morphology in the CMDs at different distances from the cluster center 
(see Fig.~\ref{6256rad}).
As shown in Fig.~\ref{6256rad}, the
reddest RGB component becomes progressively populated as the distance to the cluster
center increases, thus confirming that it is due to field stars.
As previously done for other highly contaminated clusters 
(see i.e. NGC~6304 in VOF05), 
the RGB fiducial ridge line has been derived considering only the
innermost region of the cluster, typically only the stars lying within 30" 
from the cluster center.
Since the absence of a prominent feature such as the red clumpy HB
makes difficult to derive the cluster reddening and distance, 
we used the empirical method (see FVO06)
rather than the differential one.
From the analysis of the cluster CMD we estimated the RGB slope and the observed
RGB\--Tip to be RGB$\rm _{slope}$=--0.052 and K$\rm ^{Tip}$=9.21, respectively. 
Using these values as input
parameters, the computational routine (assuming the {\it Bulge--like 
scenario}), gives a reddening E(B--V)=1.2, an intrinsic distance modulus 
(m-M)$_0$=14.79, and metallicities [Fe/H]=--1.63~dex, and [M/H]=--1.43~dex.
The distance estimate agree with the value by H96, and the
derived metallicity is consistent with those published by \citet{sf04}
 ([Fe/H]=--1.35~dex), based on low--medium resolution K spectra, and by \citet{b98}
 ([Fe/H]=--1.01~dex) from integrated optical spectroscopy.
\subsection*{NGC~6266}
NGC~6266 is a high density (log$\rho \sim$5.34, \citet{djo93}), 
moderately reddened (E(B--V)=0.44, H96; E(B--V)=0.47, S98) 
and massive (M$_V$=-9.19, H96) cluster, which has been subject 
of several dynamics, kinematics and stellar variability studies. In fact, this
cluster turns out to be particularly interesting because: {\it i)} it is the
fifth (after Ter~5, 47~Tuc, M~15 and M~28) 
ranking GC for wealth Millisecond Pulsars (MSPs), and it is the only GC
exclusively populated by MSPs in binary systems \citep{pos03}; 
{\it ii)} it hosts
more than 200 RR~Lyrae variables \citep{con05}, whose periods place 
this cluster in the Oosterhoff type~I group, though the cluster HB morphology
(a blue HB component and a very extended blue tail) is similar to 
Oosterhoff type~II GCs; {\it iii)} recent {\it Chandra} X--ray
observations have revealed a very large number of X--ray sources, suggesting
the presence of a high number of cataclysmic (and/or interacting) binaries;
{\it iv)} its radial velocity is smaller than the escape velocity from the Bulge,
indicating that, whatever its origin would be, it will never escape from the 
Bulge \citep{din03}.
\citet{giacomo06}  published the most extensive photometry based on
a combination of multi--band high--resolution HST and
wide--field ground based observations aimed at studying the cluster dynamical
state, finding that the cluster has not experienced core collapse yet. 
No IR photometric studies for this cluster have been published so far.

The observed CMD in the [K, J--K] plane (see Fig.~\ref{cmds}) shows a well
populated RGB and a blue extended HB, suggesting a low--intermediate
metallicity. As expected, in the near--IR plane
the high level of contamination by foreground disk stars affecting the optical
ground--based CMD of \citet{giacomo06} is drastically reduced, allowing us a
clear definition of the RGB ridge line from the tip down to the Sub--Giant Branch
(SGB), at K$\sim$16.
To estimate the cluster metallicity, we have adopted the distance modulus 
(m--M)$_0$=14.11 of \citet{giacomo06} (based on the \citet[][--hereafter F99]{frf99} distance scale), 
and the reddening value from H96. The global set of
photometric indices (in both [K, J--K] and [H, J--H] planes) computed for NGC~6266 have
yielded a [Fe/H]=--0.98~dex and [M/H]=--0.82~dex, in good
agreement with the values in the literature (see i.e. H96).
\subsection*{NGC~6273}
NGC~6273 is an intermediate--low metallicity ([Fe/H]=--1.68, see H96),
medium--concentration \citep[c=1.5][]{djo93} cluster, and  
it is also the second most luminous (M$_V$=-9.50) in the 
\citet{djo93} compilation.  
It has been the subject of a monographic study by \citet{pio99}, based on HST B,
V observations, aimed at understanding the nature of its extended HB blue tail,
which shows a clear double--peaked distribution and a well defined gap. From the
analysis of the CMD the authors estimated an intrinsic distance modulus
(m--M)$_0$=14.77, and an average reddening E(B--V)=0.47, though their optical
CMD is strongly affected by differential reddening ($\Delta$E(B--V)=0.2~mag). 
The only IR CMD available in the literature is the one 
published by \citet{D00}, but it is poorly populated and
quite shallow, reaching only K$\sim$15. From the [K, J--K] CMD, \citet{D00}
estimated a reddening E(B--V)=0.38, in excellent agreement with the value
listed in literature (i.e. E(B--V)=0.38, H96, and E(B--V)=0.31, 
S98).

Fig.~\ref{cmds} shows our observed CMD in the [K, J--K]
plane along with the derived RGB ridge
line (solid line). The cluster RGB is well populated even in the brightest
magnitude bin, and the HB appears as a vertical structure almost parallel to
the RGB at (J--K)$\sim$0.2, suggesting a low--intermediate metallicity.
Our photometry is deep enough to reach $\sim$1~mag below the Main Sequence--Turn Off
(MS--TO), however the SGB is quite scattered, preventing an
accurate measurement of the MS--TO luminosity, hence a reliably cluster age
estimate. Because of its HB morphology, the cluster reddening and distance have
been derived by using the empirical method of FVO06.
From the analysis of the observed CMD we derived the RGB slope (RGB$\rm _{slope}$=--0.063) and the 
the RGB tip (K$\rm ^{Tip}$=8.57). Adopting these two values as input parameters, we found 
E(B--V)=0.40, (m--M)$_0$=14.58, [Fe/H]=--1.40, and [M/H]=--1.21, for the cluster reddening, distance and
metallicity, respectively. The derived distance is consistent (within 0.2~mag) with the \citet{pio99} 
estimate
and (within 0.1~mag) with the value listed by H96. Our reddening measurement agree well
with the values in literature (see i.e. H96, S98, \citet{pio99,D00}).
\subsection*{NGC~6293}
\begin{figure}
\plotone{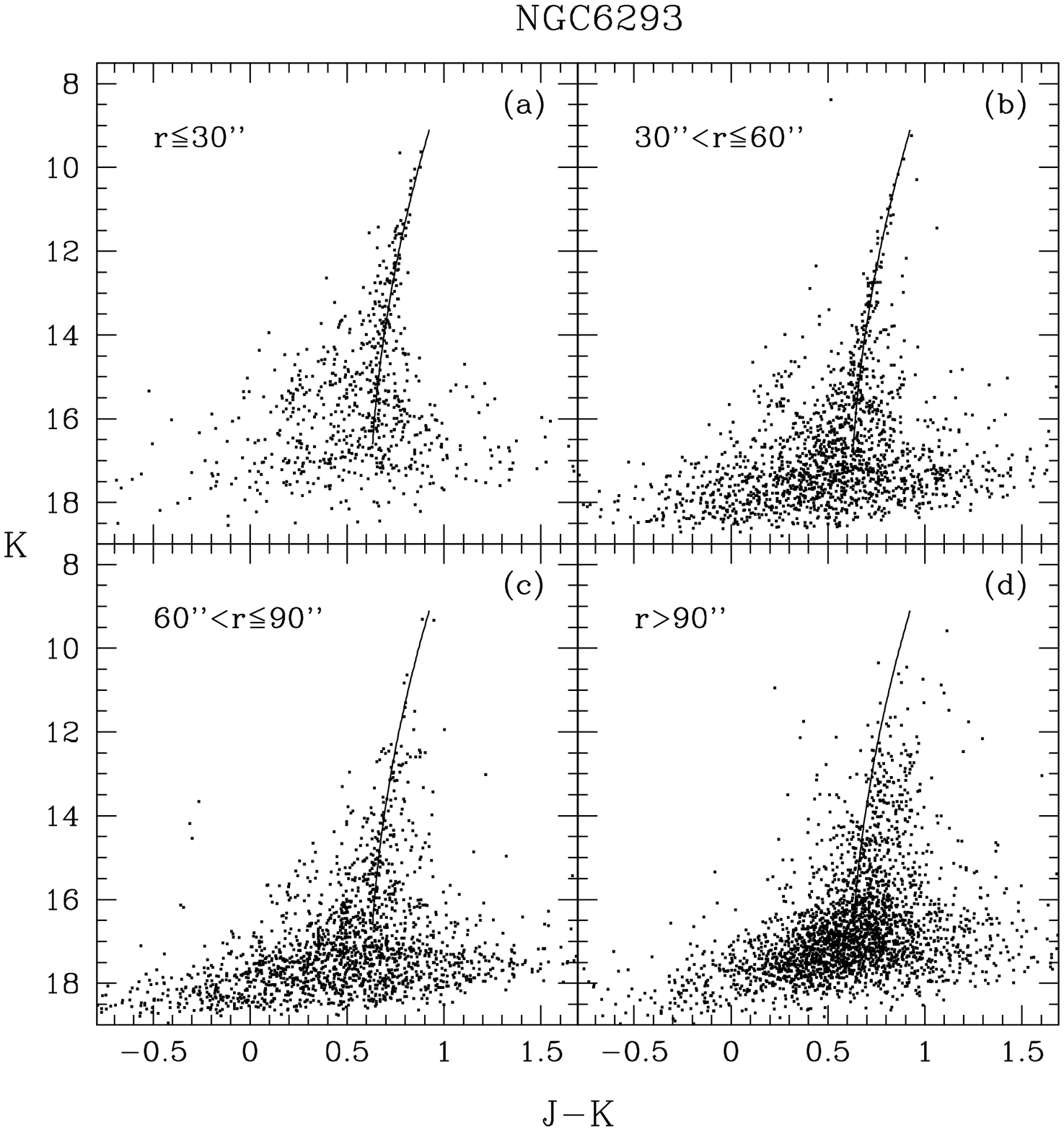}
\caption{\label{6293rad} [K, J--K] CMDs of NGC~6293 at
different distances (r) from the cluster center.}
\end{figure}

NGC~6293 is the most metal--poor cluster in the observed sample.
In the literature, the best
available photometries are those published by \citet{D00} and \citet{ptt02}.
\citet{ptt02} presented HST B, V CMD, which shows an extended blue HB, 
confirming the
low metal content of this cluster. The near--IR CMD of \citet{D00} is 
poorly populated, and only reaches
K${\sim}$15. From the IR data, the
author estimated a reddening E(B--V)=0.04, which is significantly lower than
the value listed in the literature (E(B--V)= 0.39, H96; E(B--V)= 0.60, S98). 
A careful examination of the derived CMD in Fig.~\ref{cmds}, reveals the presence 
of a quite scattered RGB, which cannot be explained in terms of
photometric errors.
In order to understand the nature of the observed scatter, as already done for NGC~6256,
in Fig.~\ref{6293rad} we compared the radial CMDs in 4 annuli at different distances
from the cluster center.  
In the
innermost 30" region (Fig.~\ref{6293rad}, {\it panel a}) 
the RGB is quite narrow,
and progressively spread out with increasing the distance from the
cluster center. A field component redder than the cluster RGB is present at 
r$>$60". This high level of field
contamination is not surprising since NGC~6293 lies in the
Ophiuchus complex, a dense stellar region towards the Galactic center. 
Hence, the RGB fiducial ridge
line has been derived using only
the stars in the innermost 30" cluster region, where the field
contamination is low. The measured reddening, distance, and metallicity estimates 
(E(B--V)=0.30, (m--M)$_0$=15.10, [Fe/H]=--1.73, and [M/H]=--1.55) have been obtained by using the
empirical method of FVO06, with the following input parameters: RGB$\rm _{slope}$=--0.048 and
K$\rm ^{Tip}$=9.24. The derived cluster parameters
nicely agree with the corresponding values
listed by H96. In particular, the 
derived photometric metallicity is consistent (within $\approx$0.2~dex) with the 
spectroscopic estimate published by \citet{lee02} (i.e. [Fe/H]=--1.99~dex).

\subsection*{NGC~6316}
Among the observed clusters, NGC~6316 is one of the most metal--rich and
with the largest distance to the Galactic center. The
[K, V--K] CMD presented by \citet{D92} shows a red clumpy HB
and a quite well populated upper RGB, the photometry reaching only 
$\sim$1~mag below the HB. They
estimated a reddening E(B--V)=0.6, which is consistent within the values listed by H96
(E(B--V)=0.55) and byS98 (E(B--V)=0.98).
\citet{ptt02} presented a HST [V, B--V] CMD of this cluster, showing a HB
morphology close to that of 47~Tuc. However the moderately high
reddening, and the contamination by foreground disk stars prevented a clear
definition
of the RGB, which appears considerably scattered even in the brightest bins.
As shown in Fig.~\ref{cmds} our CMD in the 
[H, J--H] plane is deep enough to properly sample the 
entire RGB, from the base up to the tip, thus allowing us a safe definition of
the RGB fiducial ridge line. From the CMD and the derived LF, we estimated a
reddening E(B--V)=0.56, which is intermediate between the \citet{D92} and H96 
estimates,
and an intrinsic distance modulus (m--M)$_0$=15.33, slightly longer
($\sim$0.13~mag) than the H96 value. Since no K--band 
data were taken for
this cluster, to estimate its metallicity we only measured the photometric
indices in the [H, J--H] plane. Our findings,
[Fe/H]=--0.60~dex and [M/H]=--0.45~dex, nicely agree with the value
listed by H96, confirming that NGC~6316 
has a metallicity slightly higher than 47~Tuc, as previously suggested by
\citet{D92} on the basis of the RGB slope measurement.
\subsection*{NGC~6355}
NGC~6355 lies behind a dark nebula in the eastern extension of the
Ophiuchus complex, and it is thus affected by a relatively high extinction
(i.e. E(B--V)=0.75, H96; E(B--V)=1.15, S98).
\citet{ort03} published a [V, V--I] CMD 
showing a large field contamination and a
barely detectable blue HB. From the analysis of the CMD, the authors
found E(B--V)=0.78, (m--M)$_0$=14.73 and [Fe/H]=--1.3~dex. 

\begin{figure}
\plotone{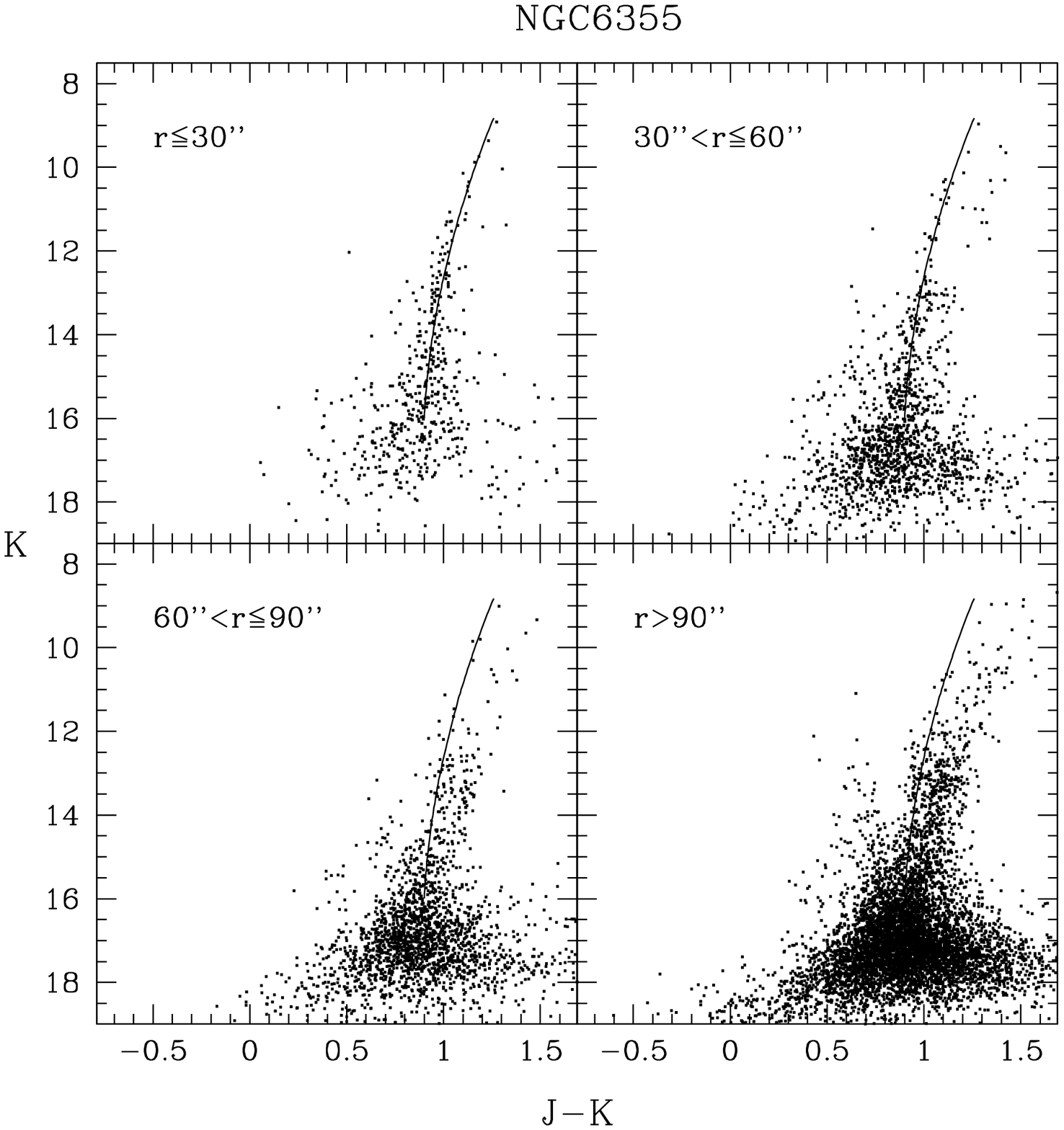}
\caption{\label{6355rad} [K, J--K] CMDs of NGC~6355 at
different distances (r) from the cluster center.}
\end{figure}

Our derived CMD, shown in
Fig.~\ref{cmds}, exhibits a blue HB at 0.2$\leq$(J--K)$\leq$0.6 and 
14.5$\leq$K$\leq$16, and a quite scattered RGB.
As expected, in the [K, J--K] and [H, J--H] planes,
the high level of contamination by disk stars strongly
affecting the \citet{ort03} CMD, is
significantly reduced. However, in our CMD the main source of  
contamination is due to the foreground Bulge field stars, which 
produce the observed RGB split. In fact, as shown in 
the radial CMDs of Fig.~\ref{6355rad}, by
increasing the distance from the cluster center, a secondary red component becomes
progressively more pronounced, causing the observed scatter in the RGB.
As done for NGC~6256 and NGC~6293, the RGB ridge line has been 
derived by using
only the stars lying within 30" from the cluster center, and 
the main cluster properties have been estimated by using the empirical method
of FVO06.
From the observed CMDs, we measured RGB$\rm_{slope}$=--0.068 and
K$\rm_{Tip}$=8.854. Adopting these two values we find 
E(B-V)=0.82, (m-M)$_0$=14.70, [Fe/H]=--1.42~dex and [M/H]=--1.22~dex for the cluster reddening,
distance and metallicity.
Our findings nicely agree with the \citet{ort03} results.

\subsection*{NGC~6388}
NGC~6388 is a moderately reddened (i.e. E(B--V)=0.37, H96;
E(B--V)=0.39, S98; E(B--V)=0.40, \citet{pri02}), metal--rich 
([Fe/H]=-0.60, H96, \citet{pri05}) cluster located in the outer
Bulge region (see Fig.~\ref{map}). Since the discovery by \citet{rmr97} of an extended
blue HB (as well as in NGC~6441) in its HST--based CMD, NGC~6388 has been subject 
of several optical photometric studies 
\citep[see i.e.][and reference therein]{moe99,ptt02,pri02}. 
NGC~6388 displays a quite peculiar HB morphology. In fact,
besides a well populated red HB clump (a feature
normally predicted by the stellar evolution theory in the case of old, metal--rich
populations) it shows
an extended blue tail and a population of RR~Lyrae variables
\citep{pri02}. NGC~6388 and NGC~6441 are thus the most metal--rich examples 
of {\it Second--parameter} affecting the HB morphology. 
From the analysis of the cluster CMD based on Washington 
photometry, \citet{hug06} claimed that NGC 6388 has a RGB too broad to be chemically
homogeneous, suggesting a metallicity spread of $\delta$[Fe/H]=$\sim$0.3~dex.
However, very recently, no intrinsic metallicity spread 
has been found by \citet{eug06}, who performed
high--resolution (R$\approx$40,000) optical spectroscopy of seven cluster members,
finding an average [Fe/H]=--0.44~dex with a {\it rms}=0.04~dex. 

No IR photometric studies have been published so far for this cluster. 
Fig.~\ref{cmds} shows the first CMD in the [K, J--K] plane, and the derived RGB
ridge line. As expected, the observed CMD is characterized by a well defined red HB
clump suggesting a high metallicity, while the extended blue HB is barely
visible as the vertical structure at 15.5$\lesssim$K$\lesssim$17.5 and
0$\lesssim$(J--K)$\lesssim$0.6. The RGB appears well populated over its entire
extension, from the base (K$\sim$17) up to the tip, allowing us 
a detailed description of
its morphological features. From the comparison of the cluster CMDs and LFs with those
of 47~Tuc, we estimated the reddening and distance to be E(B--V)=0.44 and
(m--M)$_0$=15.38, respectively. The photometric indices measured along the RGB, in the
[M$\rm _K$, (J--K)$_0$] and [M$\rm _H$, (J--H)$_0$] absolute planes, yielded the
following metallicity estimates: [Fe/H]=--0.61~dex and [M/H]=--0.42~dex, in
good agreement with previous literature values.
\subsection*{NGC~6401}
NGC~6401 is located in the inner Bulge region, at $\sim$0.8~Kpc (see Table~\ref{tab1} and
Fig.~\ref{map})
from the Galactic center. Its physical parameters are not well determined in the
literature. In particular, the extinction estimates range from E(B--V)=0.53
\citep{bar99} to E(B--V)=0.98 (S98), and the cluster distance 
values are between
(m--M)$_0$=14.39 (H96) and (m--M)$_0$=14.76 \citep{bar99}.
From the analysis of the observed [V, V--I] CMD, \citet{bar99} concluded that
NGC~6401 is metal--rich, with a metallicity close to 47~Tuc, while 
from low--resolution optical spectroscopy \citet{Mi95} found [Fe/H]=--1.1~dex.

The near--IR CMD here obtained is shown in Fig.~\ref{cmds}, along 
with
the derived RGB ridge line. The most interesting features are: {\it i)} a
blue HB tail at 0.4$\leq$(J--K)$\leq$0.9 and 13$\leq$K$\leq$16; {\it ii)} a
rather scattered RGB, which seems split in 2 separated sequences, 
particularly in the [K, J--K] plane. Note that, the presence of a blue extended HB
is confirmed by the \citet{ptt02} work based on HST BV observations.

\begin{figure}
\plotone{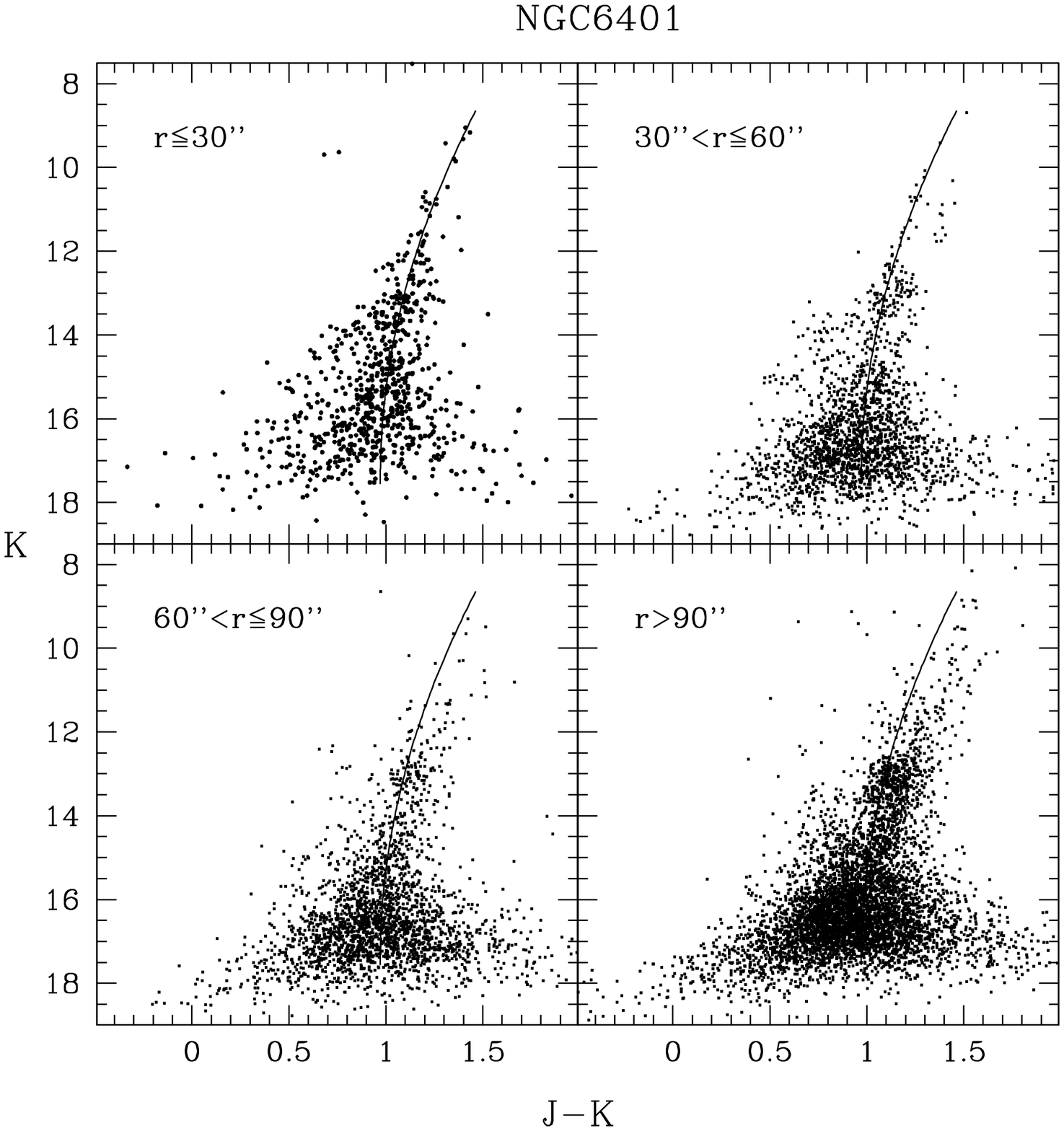}
\caption{\label{6401rad} [K, J--K] CMDs of NGC~6401 at
different distances (r) from the cluster center.}
\end{figure}

Fig.~\ref{6401rad} clearly shows that, in the
innermost region, within 30" from the cluster center, the CMD exhibits a
blue HB and a quite narrow RGB, while by increasing the distance from the
cluster center the CMD starts to be characterized
by a red clumpy HB and by a RGB significantly redder than the
cluster mean loci. Hence, the observed RGB spread is likely due to the bulk of
the Bulge field population which is metal--rich.
In order to derive the cluster main properties we used the FVO06 empirical method,
assuming the {\it Bulge--like scenario}. We measured the cluster RGB
slope considering only those stars lying in the innermost cluster region
(RGB$\rm_{slope}$=-0.065) and we estimated the observed (K=8.69) 
RGB tip by using the
brightest stars in our catalog lying along the RGB ridge line. We used
these values as input parameters for the computational routine and we obtained
the following reddening, distance and metallicity estimates: E(B--V)=1.1,
(m--M)$_0$=14.43, [Fe/H]=--1.37~dex and [M/H]=--1.20~dex. The derived photometric metallicity is
consistent with the spectroscopic result by \citet{Mi95} ($\Delta$[Fe/H]=0.25~dex)
and significantly lower than that found by
\citet{bar99} ($\Delta$[Fe/H]$\sim$0.7~dex).

\subsection*{NGC~6642}
The moderately reddened cluster NGC~6642 (i.e. E(B--V)=0.41, H96; 
E(B--V)=0.40, S98)
has been observed in the IR and in the optical by \citet{D00} and
\citet{ptt02}, respectively. 
However both the 
published CMDs show sequences which
are not well populated, and with a remarkable scatter.
Fig.~\ref{cmds} shows our derived CMD in the [H, J--H] plane, along with the
mean RGB ridge line. The main features of the CMD are a blue HB and a steep RGB,
typical of low--intermediate metallicity clusters. In this case, 
since no K photometry is available, the reddening and the
distance have been estimated by matching the cluster
sequence loci in the CMD to those of the reference cluster NGC~6752.
By using our derived
reddening (E(B--V)=0.60) and distance ((m--M)$_0$=14.68), we measured
the photometric indices, giving the following
photometric metallicity estimates: [Fe/H]=--1.20~dex and [M/H]=--1.03~dex.
These values are fully consistent (within 0.2~dex) with the
corresponding values listed by H96 ([Fe/H]=--1.35) and by  
\citet{Mi95} ([Fe/H]=--1.40), based on
low--resolution optical spectroscopy. Note that, also the obtained 
distance nicely agree with the value listed by H96 (i.e. (m--M)$_0$=14.63).
\subsection*{Terzan~3}
Terzan~3 is a low concentration cluster (c=0.70, \citet{djo93}) 
projected on the outskirts of the
Bulge (see Fig.~\ref{map}). The only 
published photometry is the one by \citet{barb98b}. From the analysis of the derived [V,
B--V] CMD, whose major features are a red HB and a moderately bent RGB,
they derived E(B--V)=0.72, (m--M)$_0$=14.05 and [Fe/H]$\sim$--0.70~dex. 

Our CMD along with the RGB fiducial ridge line is shown in Fig.~\ref{cmds}. 
This high--resolution IR
photometry reaches $\sim$2~mag below the MS--TO.
The presence of a red clumpy HB at K$\sim$13.42 suggests a moderately high metallicity,
like 47~Tuc. 
The analysis of the cluster CMDs and LFs yields a reddening E(B--V)=0.73 (in
excellent agreement with the values found by \citet{barb98} and S98), an intrinsic
distance modulus (m--M)$_0$=14.54 and metallicities [Fe/H]=--0.82~dex and
[M/H]=--0.63~dex. The significant ($\sim$0.5~mag) discrepancy between our 
and \citet{barb98b} distance is due to a different assumption for the
selective--to--total absorption coefficient (R$\rm _V$=3.1, in this study; 
R$\rm _V$=3.3 in the \citet{barb98b} work) and
for the distance scale (F99 here, and
\citet{jon92} in \citet{barb98b} paper).
\subsection*{Terzan~5}
Terzan~5 is a compact and massive cluster located in the very inner Bulge
region. In the last few years this extremely dense 
cluster has been subject of several studies aimed at investigating its stellar
population 
\citep[i.e.][]{ort96,Ort01,cohn02,origlia04}, as well as
at understanding stellar interactions and GC dynamics 
\citep[i.e.][]{heinke,scott}, having the highest stellar interaction rate of
any Galactic GC \citep{vh87}. 
In fact among GCs, Ter~5 ranks first in wealth of
MSPs, and it hosts many interacting binaries (i.e. cataclysmic variables, X--ray
binaries).
Being embedded in a heavily obscured zone, it is affected by a large
differential reddening which makes optical observations almost impossible. In
fact, the only available optical photometry is that by \citet{ort96} who
published a CMD in the [I, V--I] plane, showing largely scattered RGB and HB.
They estimated E(B--V)=2.49, (m--M)$_0$=13.74 and
a metallicity close to NGC~6553 (i.e. [Fe/H]=--0.3, \citep{origlia02}). Later
on, a
[J, J--H] CMD based on HST observations has been published by
\citet{Ort01} and \citet{cohn02} with the aim of 
deriving more accurate cluster 
parameters such as reddening, distance, and age. From the analysis of
the derived CMD, \citet{Ort01} found Ter~5 to be coeval to NGC~6528.
\citet{cohn02} obtained an IR--CMD significantly deeper ($\sim$2mag),
confirming the age found by \citet{Ort01}. 
They, also derived E(B--V)=2.16 and d$_{\odot}$=8.7~Kpc.

Fig.~\ref{cmds} shows our observed CMD in the [K, J--K] plane along with the
derived RGB ridge line.
The dominant features of the IR--CMD are: {\it i)} a red clumpy HB; and 
{\it ii)} a
well populated RGB, from the base (K$\sim$16) up to the tip (K$\sim$8.5), which
allow us a clear definition of its ridge line.
From the analysis of the CMD and LF we derived E(B--V)=2.38, and
(m--M)$_0$=13.87. The reddening value is in between the estimates provided
by \citet{ort96} and \citet{cohn02}, while our distance measurement is 
fully consistent with
the one of \citet{ort96}. The photometric indices measured along the cluster
RGB by adopting the reddening and distance quoted above, lead the
following metallicity estimates: [Fe/H]=--0.32~dex and [M/H]=--0.19~dex, thus in
excellent agreement with the high--resolution IR spectroscopic result obtained
by \citet{origlia04} ([Fe/H]=--0.21~dex).

\subsection*{Terzan~6}
Terzan~6 is another example of heavily reddened cluster in the inner Bulge
region. The only available photometric study is the one by \citet{barb97}, 
which provided a [I, V--I] CMD.
 The authors estimated the cluster reddening
(E(B--V)=2.24)
by comparing the derived CMD with that of NGC~6553, the cluster distance
((m--M)$_0$=14.23) by using the absolute V magnitude of the HB level, and from
the RGB morphology they concluded that Ter~6 shows intermediate characteristics
between 47~Tuc and NGC6528/NGC6553.

Our IR observations of Ter~6 provided the first IR--CMD (see Fig.~\ref{cmds})
showing a red HB and a curved RGB which are both high--metallicity indicators.
From the derived CMD we found 
E(B--V)=2.35 and (m--M)$_0$=14.13. Both estimates nicely agree
with the corresponding values found by \citet{barb97}.
The set of IR photometric indices measured by using
the RGB ridge lines give a cluster metallicity of
[Fe/H]=--0.62~dex and [M/H]=--0.43~dex.

\section{The final Compilation}
The compilation presented in this study is publicy available 
through the WorldWideWeb. The Website address is:\\
\begin{center}  
http://www.bo.astro.it/$\sim$GC/ir\_archive
\end{center}
For the entire cluster sample, two separate files are available: 
the first list the J, H
and K photometry for all the measured stars in each cluster, 
together with the stars 
absolute position in R.A. and Dec.; 
the second reports the observed RGB fiducial ridge line in the
[K, J--K] and [H, J--H] planes.

Table~\ref{tab1} lists the derived parameters for the 24 clusters, namely the
reddening, the distance and the metallicity in both the adopted scales (columns
[9,10,11,12], respectively). Moreover, by
using our distance determinations, the {\it (l, b)} Galactic coordinates from H96, and
assuming a distance R$_0$=8~Kpc to the Galactic center \citep{eis03},
we also provided new estimates of the clusters distance from the sun (d$_{\odot}$), from the Galactic
center (R$\rm _{GC}$), and
the distance components X, Y, Z, in a Sun--centered coordinate system (columns
[4,5,6,7,8], respectively). Finally for each cluster, column [13] lists
the references for the IR photometry.
It is worth mentioning that Table~\ref{tab1} lists the largest, homogeneous compilation of
Bulge GCs obtained so far, whose properties have been derived in a self--consistent way.
In fact, the derived cluster reddening, distance and metallicity are
based on {\it i)} a homogeneous photometric database analyzed by using the same data
reduction procedures, and calibrated onto the 2MASS photometric system; {\it ii)} 
the F99 distance scale; 
{\it iii)} a uniform and high--resolution metallicity scale \citep{cg97}.

A summary table with the derived parameters for the global sample of clusters,
here split in Tables~\ref{tab1} and \ref{tab2}, is also provided 
in a easily machine--readable format.

\subsection{The RGB Bump and Tip}
The most interesting evolutionary features along the RGB are the so--called bump and tip.
The former flags the point (during the post--MS evolution of low mass stars) when the
narrow hydrogen--burning shell reaches the discontinuity in the hydrogen distribution
profile, generated by the previous innermost penetration of the convective envelope. 
Besides providing an obvious check on the accuracy of theoretical models of stellar
evolution, the identification of the RGB bump in stellar systems can be used as useful tool
for providing observational constraints on a number of population parameters, since it is
a sensitive function of metal content, helium abundance and stellar population age.
From the observational point of view,
as emphasized by \citet{fusi90} and F99, the combined use of the differential and integrated
LFs is the best tool to properly detect the bump.

The evolution along the RGB ends at the so--called RGB tip with helium ignition in the
stellar core. In GC stars, this event is moderately violent because it takes place in a
electron--degenerate core. Because the RGB reaches its maximum extension in luminosity for
stellar populations older than $\tau \approx$1--2~Gyr (i.e. when stars with M$\rm \le 2.0
M_{\odot}$ are evolving off the MS) 
and it remains approximately constant with increasing the age of the
population, the tip is now widely used as a standard candle for 
distance determination of stellar systems.

The dependence of the RGB bump and tip 
luminosities (in the J, H and K bands. as well as in bolometric) 
on the cluster metallicity has been investigated in VFO04b.
Here we have extended the same study to the new sample of Bulge clusters. 
In doing this, we have adopted the same
strategy followed in that paper, where the reader can find a detailed description of the 
adopted procedures. 

For the entire sample, the measurements of the main evolutionary features are
listed in Table~\ref{tab2}: the observed J, H, and K magnitude of the red HB
clump (columns~[4, 5, 6]), 
the J, H, K and bolometric bump (columns [7, 8, 9, 13], respectively), and tip 
(columns [10, 11, 12, 14], respectively) magnitudes.
The bolometric bump and tip magnitude have been obtained by using the 
bolometric corrections for Population~II giants computed by \citet{paolo98}.

\begin{figure}
\plotone{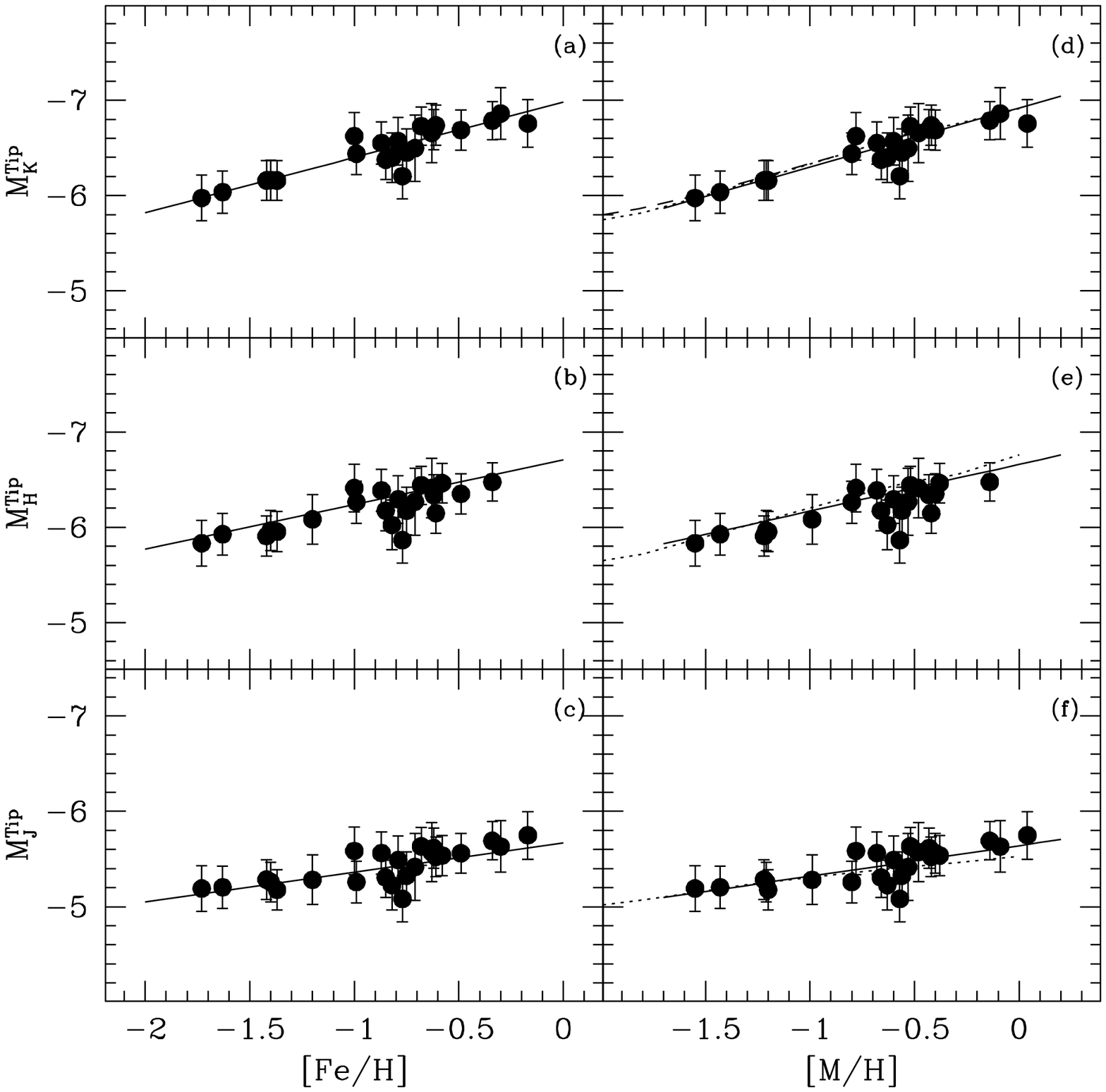}
\caption{\label{tip} The J, H and K absolute magnitudes of the RGB tip as a function
of [Fe/H] (panels a, b, c) and [M/H] (panels d, e, f) metallicities
for the clusters sample. The solid lines are the empirical relations from
VFO04b,
the dashed line is the theoretical prediction by \citet{scl97}, and the dotted lines
are the \citet{cas00} models.}
\end{figure}

\begin{figure}
\plotone{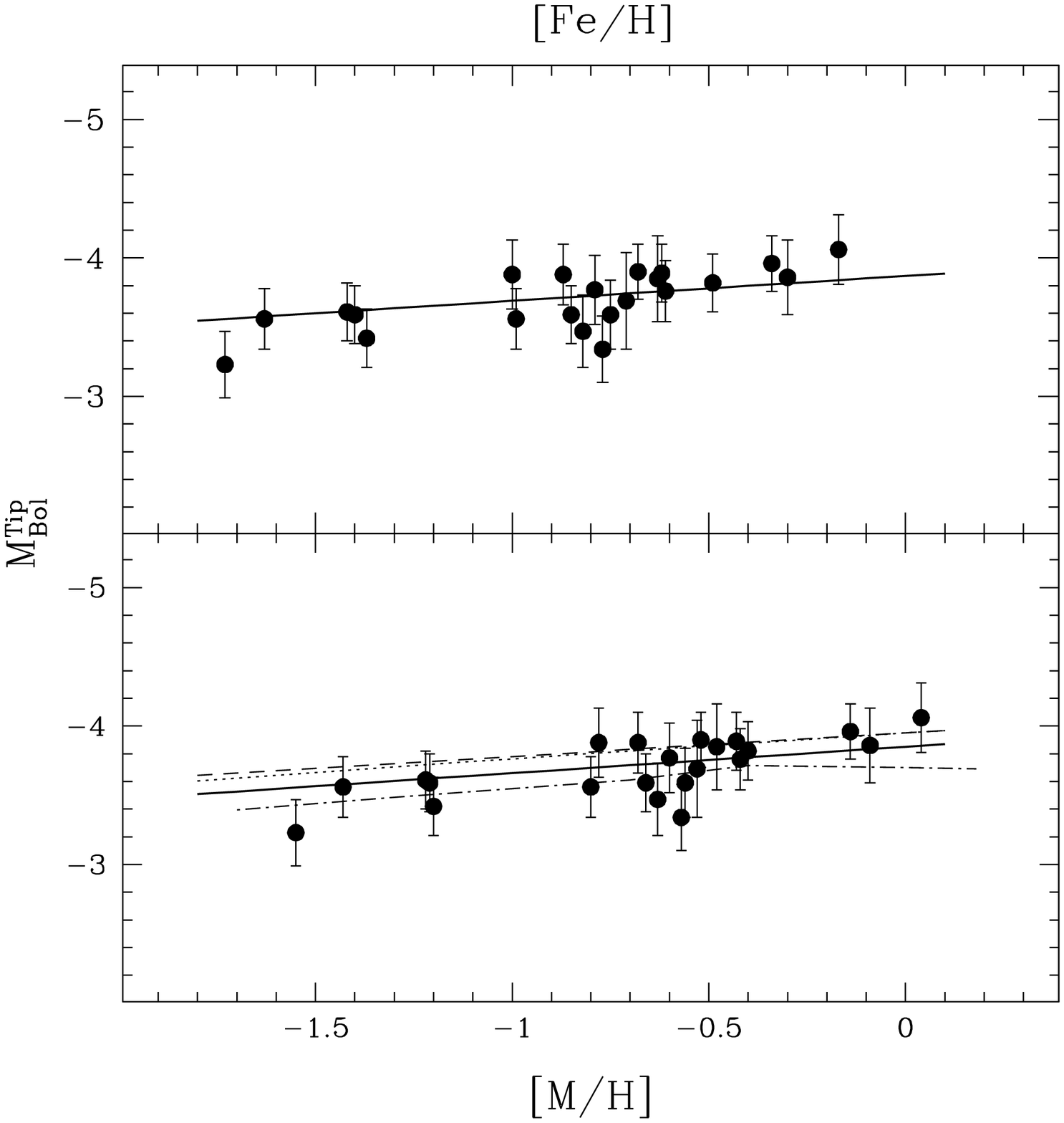}
\caption{\label{tip_bol} Bolometric magnitudes of the RGB Tip as a function of the
cluster [Fe/H] (top panel),
and [M/H] (bottom panel) metallicities. The solid lines are the empirical relation from
VFO04b. Three theoretical predictions have been plotted in the bottom panel:
\citet{cal97} (dashed line), \citet{sal97} (dotted line), and \citet{gir00}
(dotted--dashed line).}
\end{figure}

Fig.~\ref{tip} shows the absolute RGB tip magnitudes as a function of the metallicity, 
in both the adopted scales. The clusters tip determination 
nicely agree with 
the empirical relations by VFO04b (solid lines), and with the theoretical 
prediction by \citet{scl97,cas00} (dashed and dotted line, respectively). 
As shown in Fig.~\ref{tip_bol}, 
a good agreement is also found when we compare the bolometric tip magnitudes, as a
function of the clusters metallicity, with the VFO04b empirical relation  
(solid lines), and with four different theoretical expectations by 
\citet{cal97,sal97,gir00} (dashed, dotted and dotted--dashed lines, respectively).

\begin{figure}
\plotone{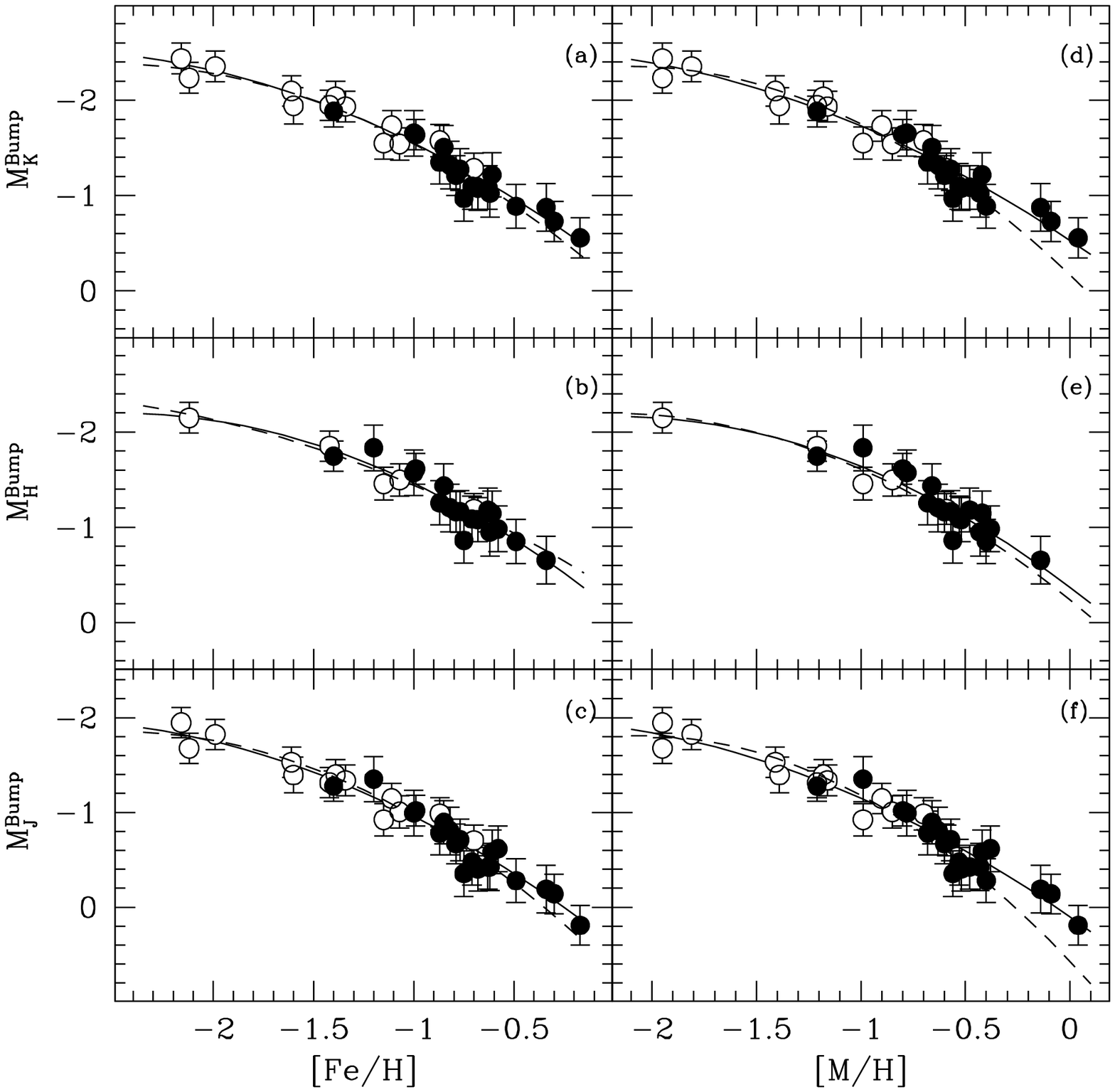}
\caption{\label{bump} The J, H and K absolute magnitudes of the RGB bump as a function
of the cluster [Fe/H] (panels a, b, c) and [M/H] (panels d, e, f) metallicities.
Filled circles, programme bulge clusters; empty circles, halo clusters presented
in VFO04b. The solid lines are the new best\--fitting relations. The dashed lines are the VFO04b
calibrations.}
\end{figure}

Fig.~\ref{bump} plots
the absolute J, H and K bump magnitudes of the bulge (filled
circles) and Halo (open circles) clusters (the latter presented in VFO04b)
{\it vs} metallicity, showing that the bump becomes rapidly fainter by increasing
the latter. Hence, since with
the present study we both enlarged the sample of high metallicity clusters and
adopted the recent [Fe/H] estimates by \citet{origlia02,origlia05}
based on high--resolution IR spectroscopy, for the two most metal--rich objects
in our sample (namely, NGC~6528 and NGC~6553), we calibrated new relations. 
As shown in Fig.~\ref{bump} the new calibrations (solid lines) differ from
the VFO04b ones (dashed lines) only in the metal--rich tail, making the bump somewhat 
brighter. 
The new calibrations of the bump J, H, K absolute magnitude, in both the
adopted metallicity scales, are as follows:
\begin{equation}
\rm M_J^{Bump}=0.39+1.63[Fe/H]+0.28[Fe/H]^2
\end{equation}
\begin{equation}
\rm M_H^{Bump}=-0.12+1.68[Fe/H]+0.34[Fe/H]^2
\end{equation}
\begin{equation}
\rm M_K^{Bump}=-0.25+1.57[Fe/H]+0.27[Fe/H]^2
\end{equation}
\begin{equation}
\rm M_J^{Bump}=0.10+1.53[M/H]+0.28[M/H]^2
\end{equation}
\begin{equation}
\rm M_H^{Bump}=-0.37+1.65[M/H]+0.38[M/H]^2
\end{equation}
\begin{equation}
\rm M_K^{Bump}=-0.53+1.47[M/H]+0.27[M/H]^2
\end{equation}

\begin{figure} 
\plotone{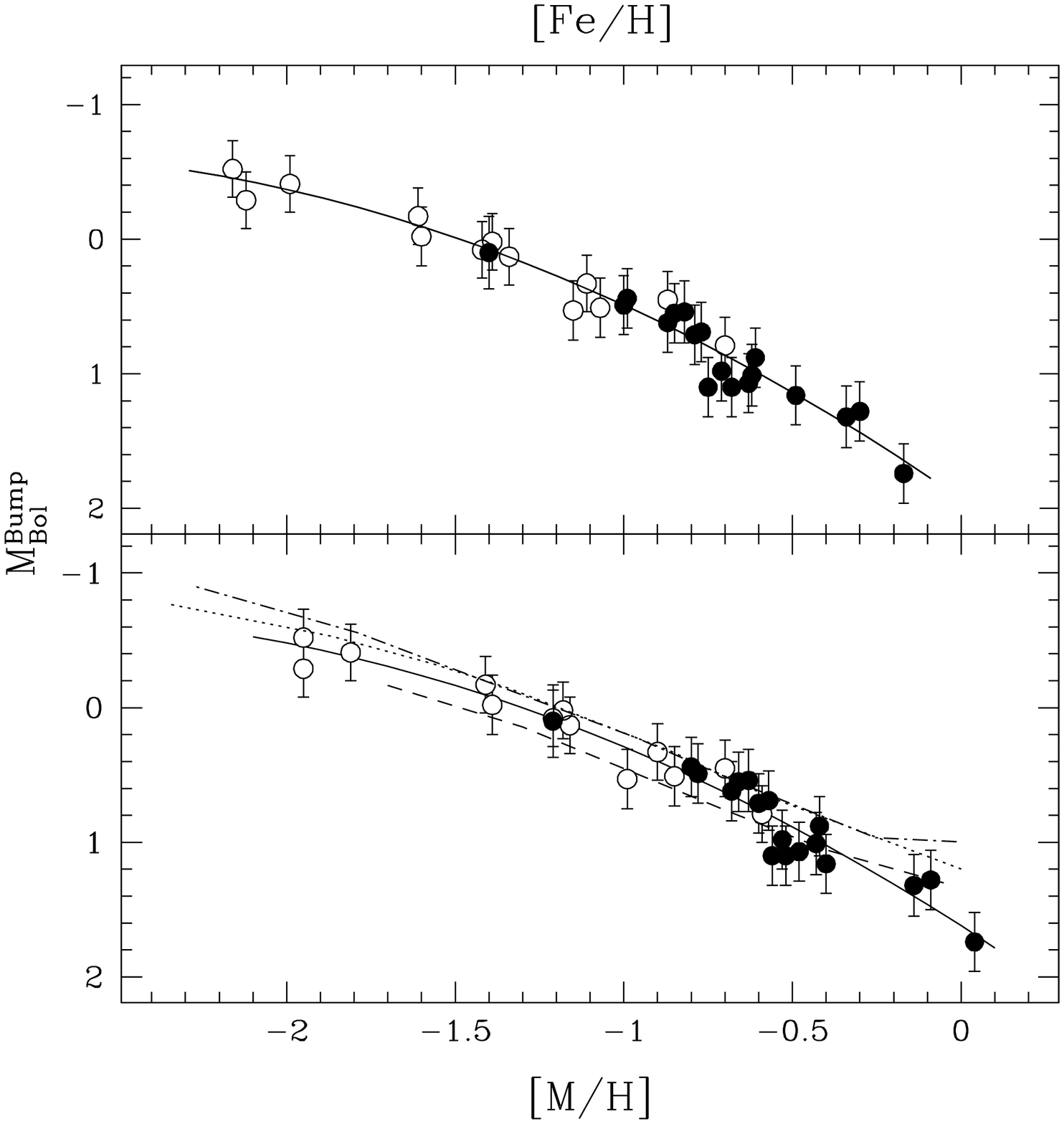}
\caption{\label{bump_bol} Bolometric magnitudes of the RGB Bump as a function of the
cluster [Fe/H] (upper panel) and [M/H] (lower panel) metallicities. 
Filled circles, programme bulge clusters; 
empty circles, halo clusters presented in VFO04b.
The solid lines are our best\--fitting relations.
Three theoretical predictions have been plotted in the lower panel:
\citet{scl97} (dotted line), \citet{gir00} (dashed line), and \citet{pie04}
(dashed\--dotted line).}
\end{figure}

Fig.~\ref{bump_bol} shows the bolometric magnitude of the bump as a function
of the cluster metallicity, together with the new best--fit relations:
\begin{equation}
\rm M_{Bol}^{Bump}=1.93+1.73[Fe/H]+0.29[Fe/H]^2
\end{equation}
\begin{equation}
\rm M_{Bol}^{Bump}=1.62+1.61[M/H]+0.28[M/H]^2
\end{equation}

The comparison between the data and the theoretical models by
\citet{scl97,gir00,pie04} (see lower panel of Fig.~\ref{bump_bol})
shows an overall good agreement.

\acknowledgements

Part of the data analysis has been performed with the software developed by P.
Montegriffo at the Osservatorio Astronomico di Bologna (INAF--AOB).

The MIUR (Ministero dell'Istruzione, dell'Universit\'a e della Ricerca) is kindly
acknowledged for the financial support.

We warmly thanks the ESO\--La Silla Observatory Staff for assistance 
during the observations.

This publication makes use of data products from the Two Micron All Sky Survey,
which is a joint project of the University of Massachusetts and Infrared
Processing and Analysis Center/California Institute of Technology, founded by
the National Aeronautics and Space Administration and the National Science
Foundation.



{\small
\begin{deluxetable}{lcccccccccccl}
\rotate
\tabletypesize{\footnotesize}
\tablecaption{\label{tab1}Coordinates, distance, reddening and metallicity for the global
sample of Bulge clusters.}
\tablehead{
\colhead{Name} & \colhead{l$^{\circ}$} & \colhead{b$^{\circ}$} & \colhead{d$_{\odot}$} &
\colhead{R$\rm _{GC}$} & \colhead{X} & \colhead{Y} &\colhead{Z}&
\colhead{E(B--V)} & \colhead{(m--M)$_0$} & 
\colhead{[Fe/H]} &\colhead{[M/H]} &\colhead{Ref.} \\
\colhead{ } & \colhead{ } & \colhead{ } &\colhead{(Kpc)} &
\colhead{(Kpc)} & \colhead{(Kpc)} & \colhead{(Kpc)} &\colhead{(Kpc)} &
\colhead{} & \colhead{} &\colhead{}&\colhead{} & \colhead{} 
}
\startdata
6256$^*$ &-12.21 & 3.31&  9.1 & 2.2&  8.9& -1.9&  0.5&  1.20&  14.79&  -1.63&  -1.43&tw\\
6266	 &-6.42 & 7.32 &  6.6 & 1.8&  6.5& -0.7&  0.8&  0.47&  14.11&  -0.99&  -0.80&tw\\
6273$^*$ &-3.13 & 9.38 &  8.2 & 1.4&  8.1& -0.4&  1.3&  0.40&  14.58&  -1.40&  -1.21&tw\\
6293$^*$ &-2.38 & 7.83 & 10.5 & 2.8& 10.4& -0.4&  1.3&  0.30&  15.10&  -1.73&  -1.55&tw\\
6304	 &-4.17 & 5.38 &  6.0 & 2.2&  5.9& -0.4&  0.6&  0.58&  13.88&  -0.75&  -0.56&VOF05\\
6316	 &-2.82 & 5.76 & 11.6 & 3.8& 11.6& -0.6&  1.2&  0.56&  15.33&  -0.58&  -0.38&tw\\
6342     & 4.90 & 9.73 &  8.4 & 1.6&  8.3&  0.7&  1.4&  0.57&  14.63&  -0.71&  -0.53&VFO04a\\
6355$^*$ &-0.42 & 5.43 &  9.0 & 1.1&  8.7& -0.1&  0.8&  0.81&  14.77&  -1.42&  -1.22&tw\\
6380     &-9.82 &-3.42 &  9.2 & 1.9&  9.0& -1.6& -0.5&  1.29&  14.81&  -0.87&  -0.68&VFO04a\\
6388     &-14.44&-6.74 & 11.9 & 4.8& 11.5& -3.0& -1.4&  0.44&  15.38&  -0.61&  -0.42&tw\\
6401$^*$ & 3.45 & 3.98 &  7.7 & 0.8&  7.7&  0.5&  0.5&  1.10&  14.43&  -1.37&  -1.20&tw\\
6440     & 7.73 & 3.80 &  8.2 & 1.2&  8.1&  1.1&  0.5&  1.15&  14.58&  -0.49&  -0.40&VFO04a\\
6441     &-6.47 &-5.01 & 13.5 & 5.7& 13.4& -1.5& -1.2&  0.52&  15.65&  -0.68&  -0.52&VFO04a\\
6528     & 1.14 &-4.17 &  7.5 & 0.8&  7.5&  0.1& -0.5&  0.62&  14.37&  -0.17&  +0.04&F00\\
6539	 &20.80 & 6.78 &  8.4 & 3.1&  7.8&  3.0&  1.0&  1.08&  14.63&  -0.79&  -0.60&O05\\
6553     & 5.25 &-3.02 &  4.9 & 3.2&  4.9&  0.4& -0.3&  0.84&  13.46&  -0.30&  -0.09&F00\\
6569	 &0.48  &-6.68 & 12.0 & 4.2& 11.9&  0.1& -1.4&  0.49&  15.40&  -0.85&  -0.66&VOF05\\
6624     & 2.79 &-7.91 &  8.4 & 1.3&  8.3&  0.4& -1.2&  0.28&  14.63&  -0.63&  -0.48&VFO04a\\ 
6637	 &1.70  &-10.30&  9.4 & 2.1&  9.3&  1.3& -1.2&  0.14&  14.87&  -0.77&  -0.57&VOF05\\
6638	 &7.90  &-7.15 & 10.3 & 2.9& 10.2&  0.3& -1.8&  0.43&  15.07&  -1.00&  -0.78&VOF05\\
6642	 &9.81  & -6.44&  8.6 & 1.8&  8.4&  1.5& -1.0&  0.60&  14.68&  -1.20&  -0.99&tw\\
Ter~3    &-14.92&9.19  &  8.1 & 2.4&  7.7& -2.1&  1.3&  0.73&  14.54&  -0.82&  -0.63&tw\\
Ter~5    & 3.81 & 1.67 &  5.9 & 2.1&  5.9&  0.4&  0.2&  2.38&  13.87&  -0.34&  -0.14&tw\\
Ter~6    &-1.43 & -2.16&  6.7 & 1.3&  6.7& -0.2& -0.3&  2.35&  14.13&  -0.62&  -0.43&tw\\
 & & & & & & & & & & & &\\ 	
\multicolumn{13}{l}{$^*$ For these clusters the metallicity estimates ([M/H]) have
been derived by using the FVO06 empirical method.}\\
\multicolumn{13}{l}{Photometry References: tw \-- this work}\\
\enddata
\end{deluxetable}
}
{\footnotesize
\begin{deluxetable}{lccccccccccccc}
\rotate
\tablecaption{\label{tab2}Observed and bolometric 
magnitude of the HB red clump, RGB bump and tip for the global sample of bulge clusters.}
\tablehead{
\colhead{Name} & \colhead{[Fe/H]} &\colhead{[M/H]} &\colhead{J$\rm ^{RC}$}&\colhead{H$\rm ^{RC}$}&
\colhead{K$\rm ^{RC}$}&
\colhead{J$\rm ^{Bump}$} &\colhead{H$\rm ^{Bump}$} &\colhead{K$\rm ^{Bump}$} &
\colhead{J$\rm ^{Tip}$} &\colhead{H$\rm ^{Tip}$} &\colhead{K$\rm ^{Tip}$} &
\colhead{M$\rm _{Bol}^{Bump}$} &\colhead{M$\rm _{Bol}^{Tip}$} 
}
\startdata
6256 & -1.63&  -1.43&  ---  &  ---  & ---  & ---  & ---  & ---  &10.63 &9.51 &9.21 & --- &-3.56\\	
6266 & -0.99&  -0.80&  ---  &  ---  & ---  &13.50 &12.75 &12.65 &9.26  &8.10 &7.85 &0.44 &-3.56\\
6273 & -1.40&  -1.21&  ---  &  ---  & ---  &13.65 &13.05 &12.85 &9.67  &8.83 &8.57 &0.10 &-3.59\\
6293 & -1.73&  -1.55&  ---  &  ---  & ---  & ---  & ---  & ---  &10.17 &9.43 &9.24 & --- &-3.23\\
6304 & -0.75&  -0.56& 13.55 & 12.85 &12.70 &14.03 &13.33 &13.13 &9.06  &8.02 &7.65 &1.10 &-3.59\\ 
6316 & -0.58&  -0.38& 14.93 & 14.25 & ---  &15.20 &14.65 & ---  &10.28 &9.17 & --- & --- & --- \\
6342 & -0.71&  -0.53& 14.25 & 13.60 &13.40 &14.65 &13.85 &13.75 &9.71  &8.67 &8.35 &0.98 &-3.69\\
6355 & -1.42&  -1.22&  ---  &  ---  & ---  & ---  & ---  & ---  &10.19 &9.30 &8.92 & --- &-3.61\\
6380 & -0.87&  -0.68& 14.95 & 14.15 &13.85 &15.15 &14.25 &13.95 &10.37 &9.12 &8.75 &0.62 &-3.88\\
6388 & -0.61&  -0.42& 14.90 & 14.27 &14.17 &15.18 &14.47 &14.33 &10.24 &9.47 &8.81 &0.88 &-3.76\\
6401 & -1.37&  -1.20&  ---  &  ---  & ---  & ---  &---   & ---  &10.21 &9.07 &8.69 & --- &-3.42\\
6440 & -0.49&  -0.40& 14.75 & 13.80 &13.60 &15.30 &14.35 &14.13 &10.02 &8.85 &8.33 &1.16 &-3.82\\
6441 & -0.68&  -0.52& 15.20 & 14.55 &14.40 &15.70 &14.85 &14.77 &10.47 &9.49 &9.12 &1.10 &-3.90\\
6528 & -0.17&  +0.04& 14.15 &  ---  &13.35 &15.10 & ---  &14.05 &9.16  & --- &7.85 &1.74 &-4.06\\
6539 & -0.79&  -0.60& 14.65 & 13.85 &13.65 &14.90 &14.05 &13.83 &10.08 &8.92 &8.47 &0.71 &-3.77\\
6553 & -0.30&  -0.09& 13.35 &  ---  &12.40 &14.05 & ---  &13.05 & 8.56 & --- &6.92 &1.28 &-3.86\\ 
6569 & -0.85&  -0.66& 14.95 & 14.40 &14.25 &14.93 &14.23 &14.08 &10.52 &9.49 &9.21 &0.55 &-3.59\\ 
6624 & -0.63&  -0.48& 13.95 & 13.40 &13.25 &14.45 &13.60 &13.65 &9.30  &8.37 &8.08 &1.07 &-3.85\\
6637 & -0.77&  -0.57& 14.05 & 13.65 &13.55 &14.28 &13.78 &13.65 &9.91  &9.08 &8.72 &0.69 &-3.34\\
6638 & -1.00&  -0.78& 14.65 & 14.05 &13.85 &14.45 &13.73 &13.58 &9.86  &8.89 &8.61 &0.49 &-3.88\\
6642 & -1.20&  -0.99&  ---  &  ---  & ---  &13.85 &13.17 & ---  &9.92  &8.92 & --- & --- & --- \\
Ter3 & -0.82&  -0.63& 14.27 & 13.57 &13.42 &14.37 &13.73 &13.50 &9.95  &8.91 &8.42 &0.54 &-3.47\\
Ter5 & -0.34&  -0.14& 15.15 & 13.80 &13.35 &15.75 &14.50 &13.90 &10.25 &8.68 &7.99 &1.32 &-3.96\\
Ter6 & -0.62&  -0.43& 15.10 & 14.07 &13.48 &15.75 &14.45 &14.00 &10.56 &9.06 &8.33 &1.01 &-3.89\\
\enddata
\end{deluxetable}
}


\begin{figure}
\plottwo{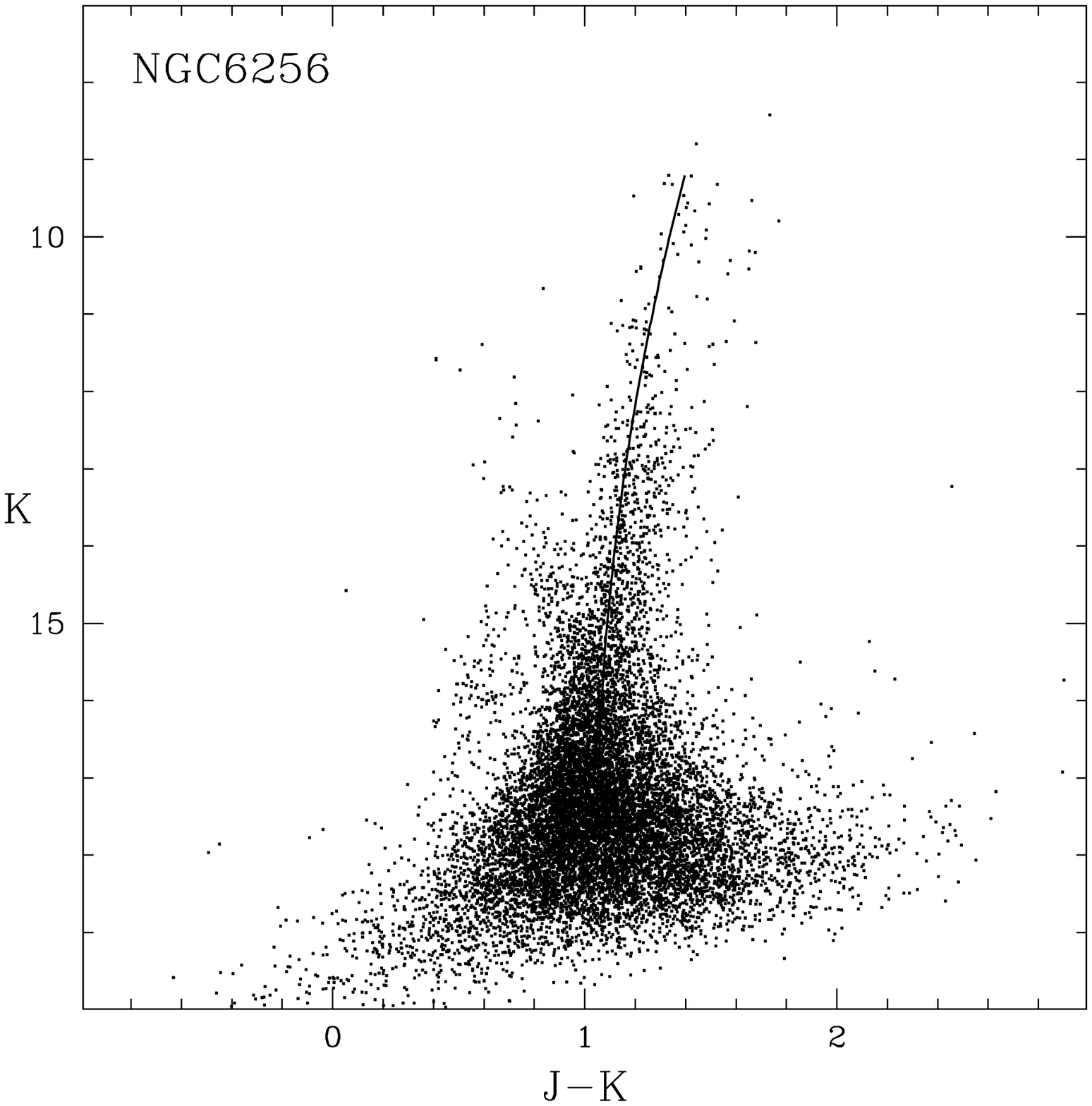}{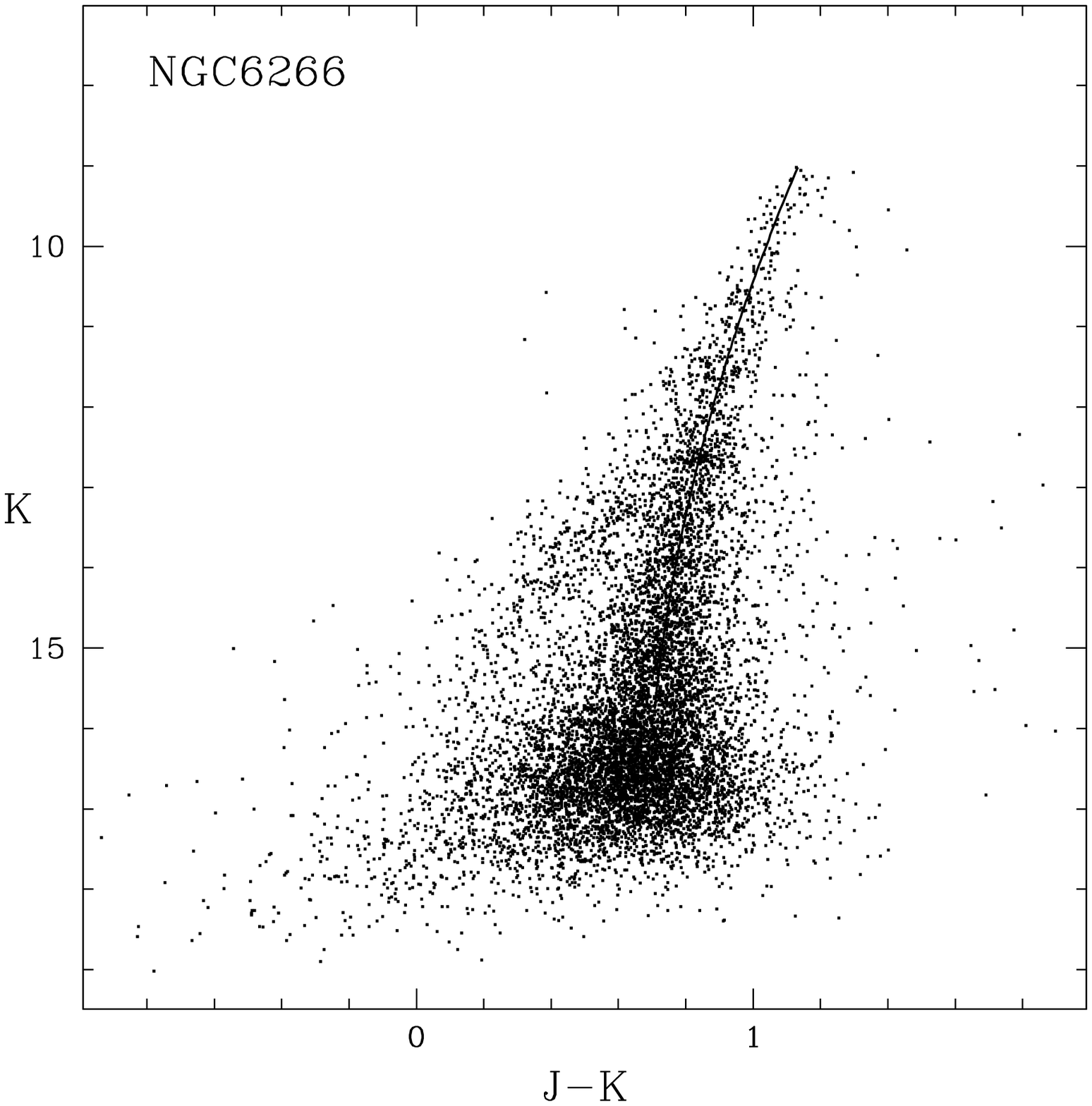}
\plottwo{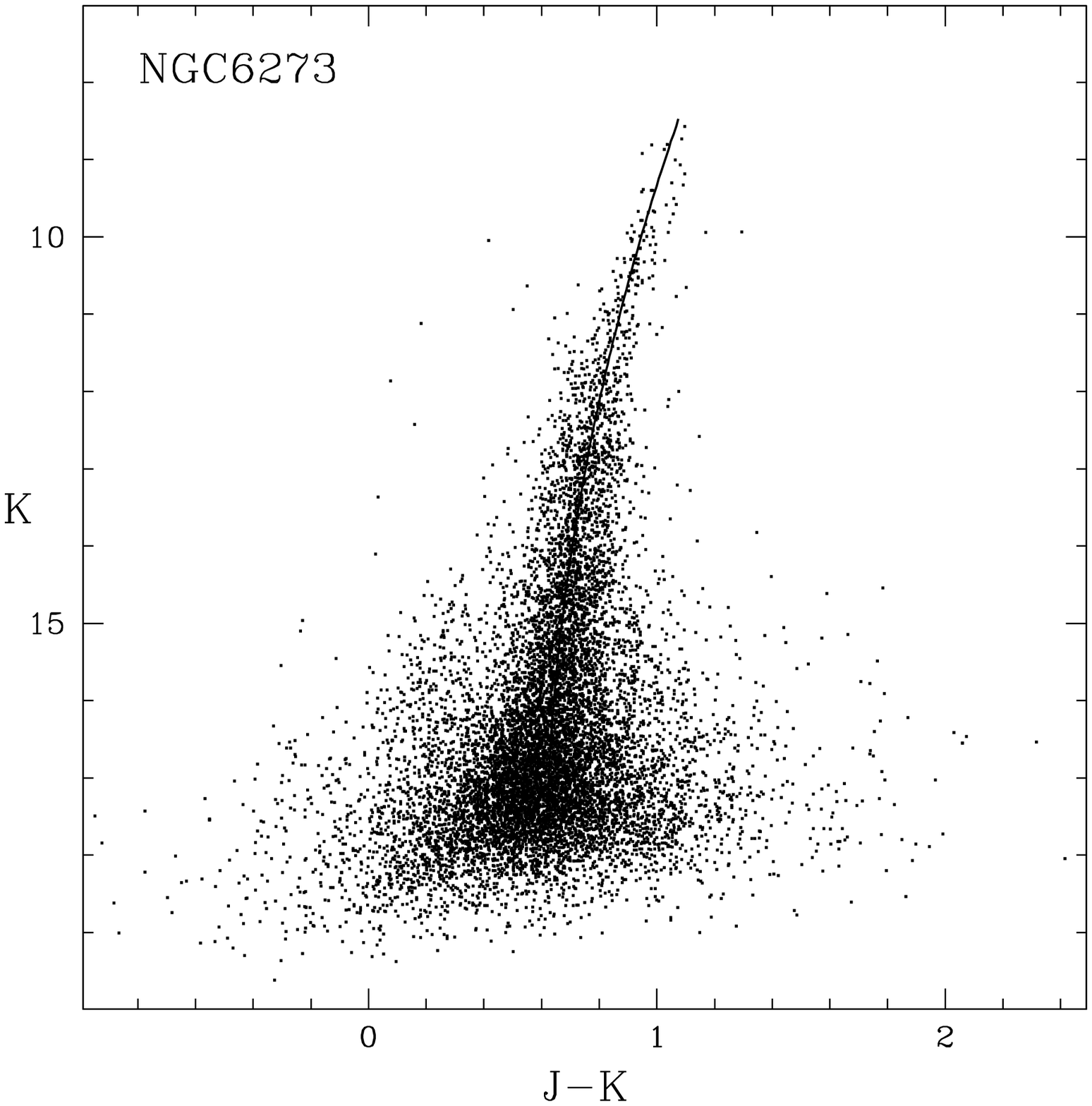}{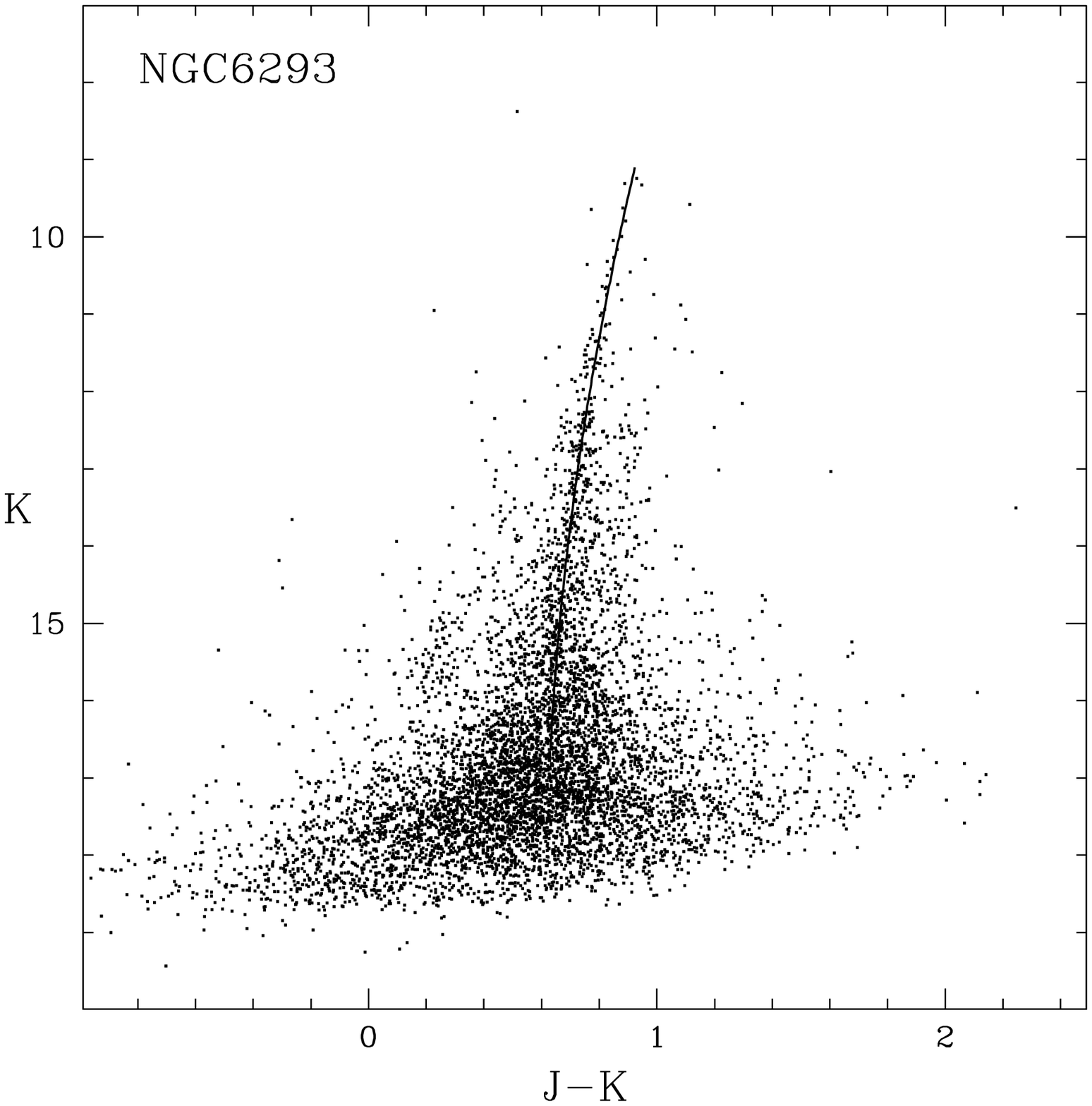}
\plottwo{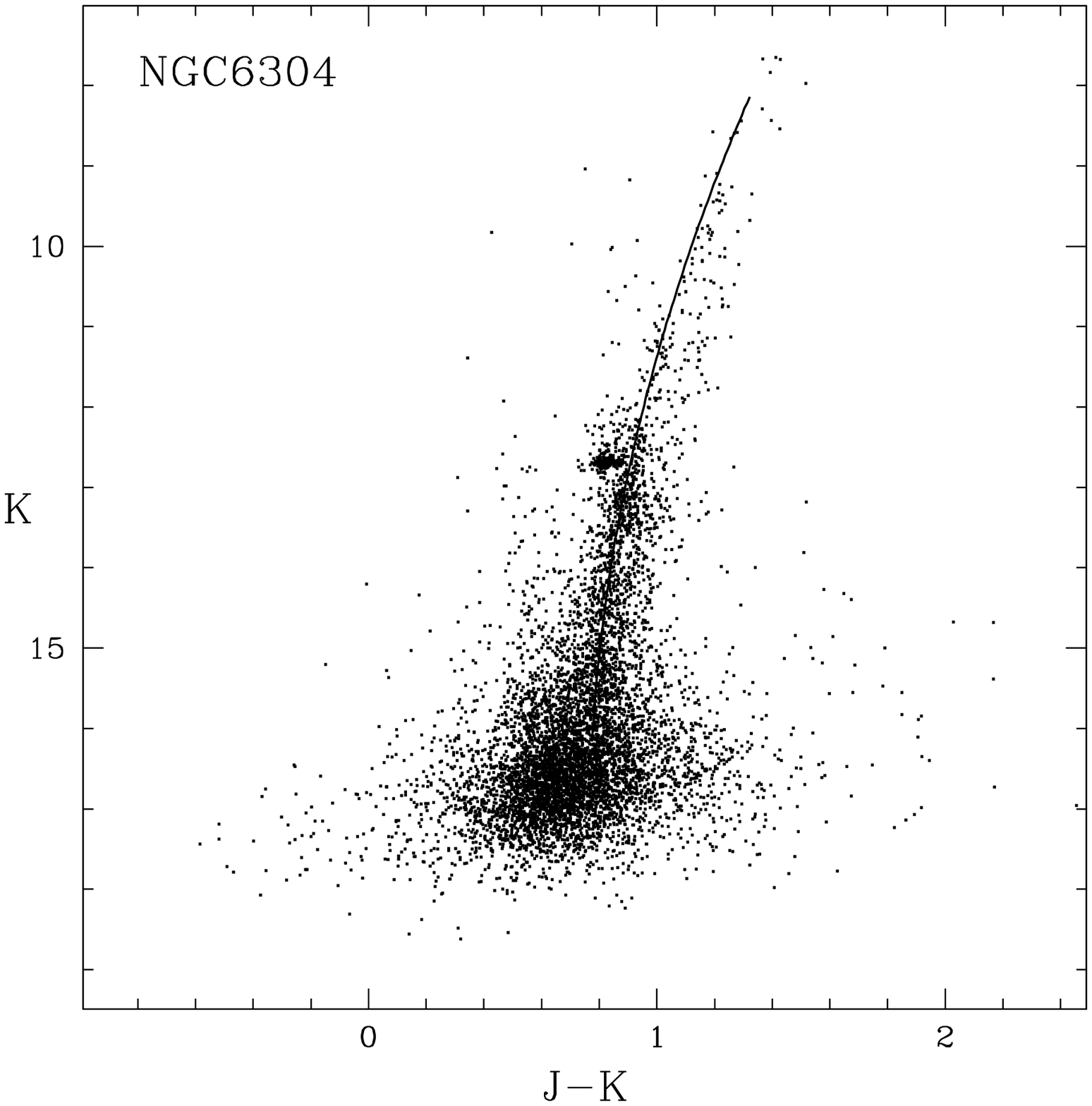}{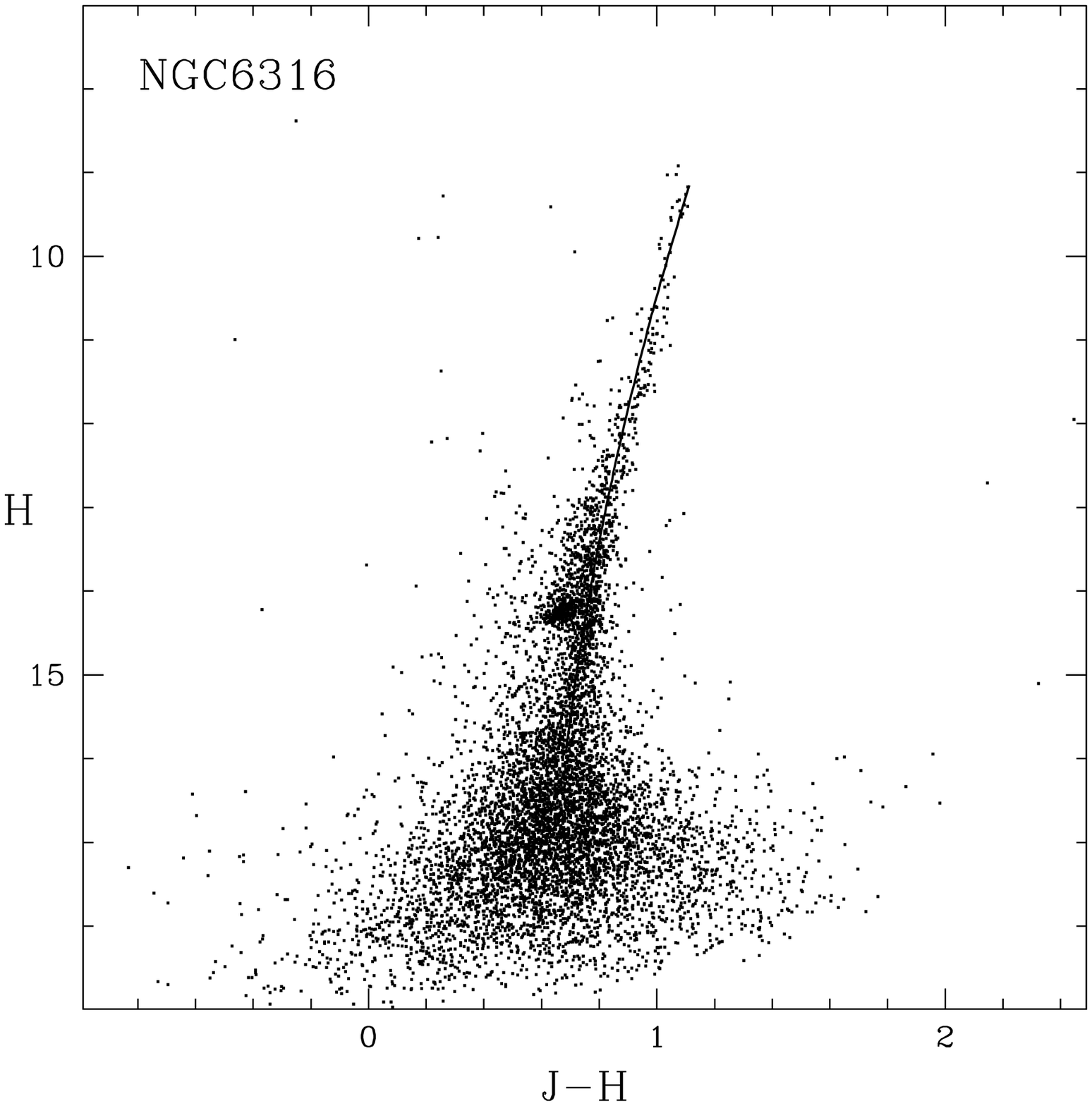}
\caption{\label{cmds} Observed near--IR CMDs and derived RGB ridge lines for the global
cluster sample.}
\end{figure}

\begin{figure}
\setcounter{figure}{1}
\plottwo{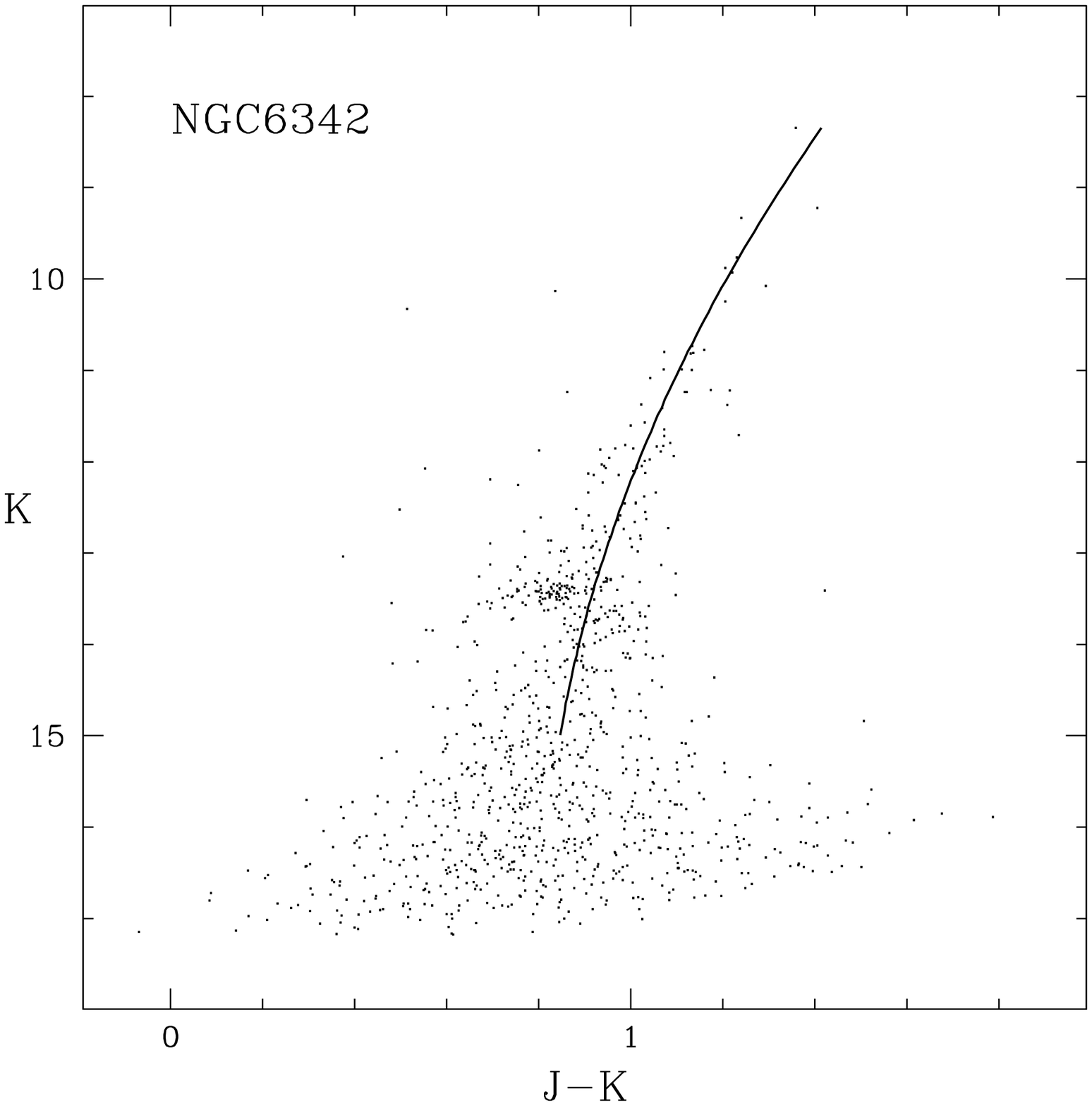}{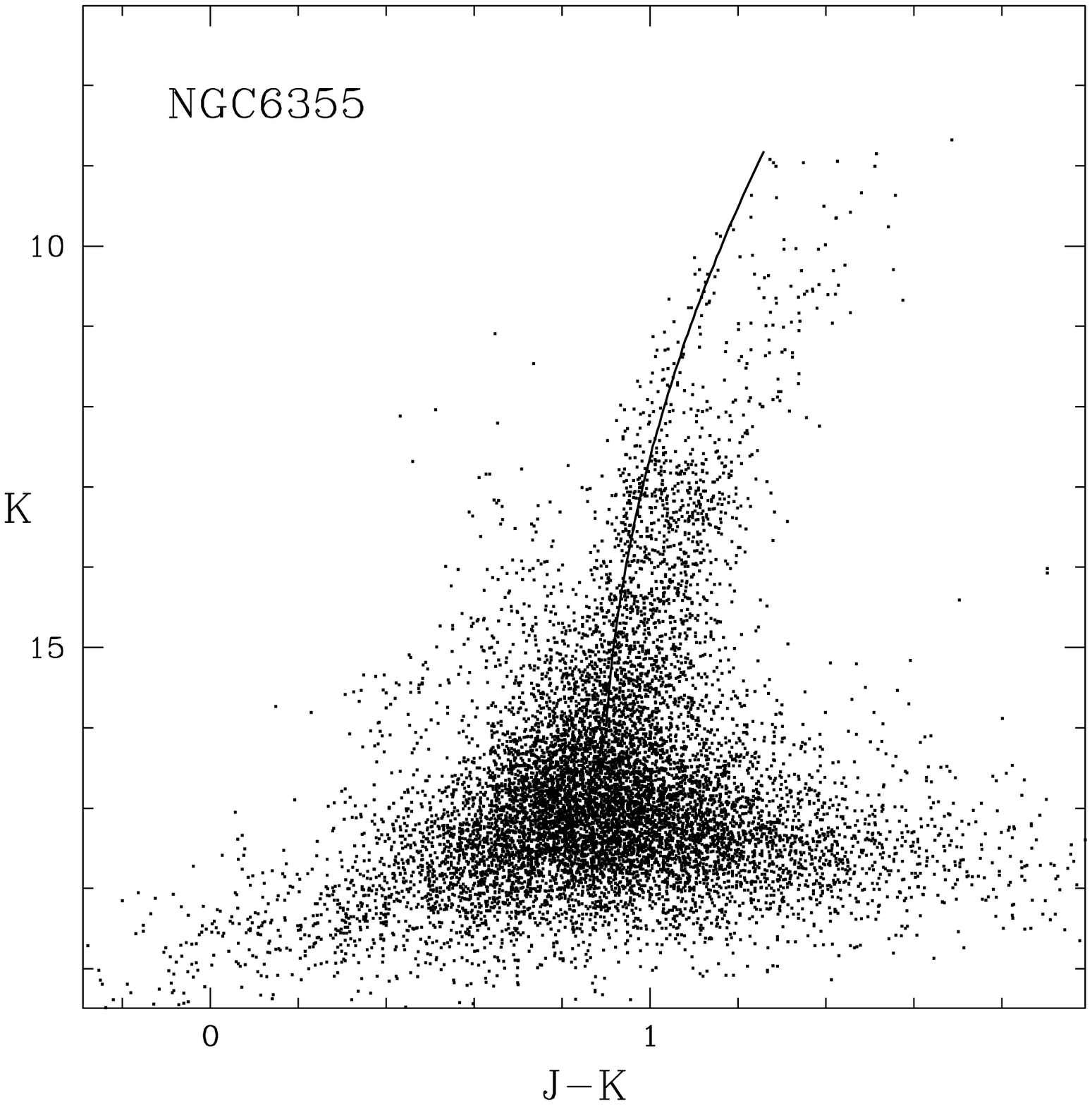}
\plottwo{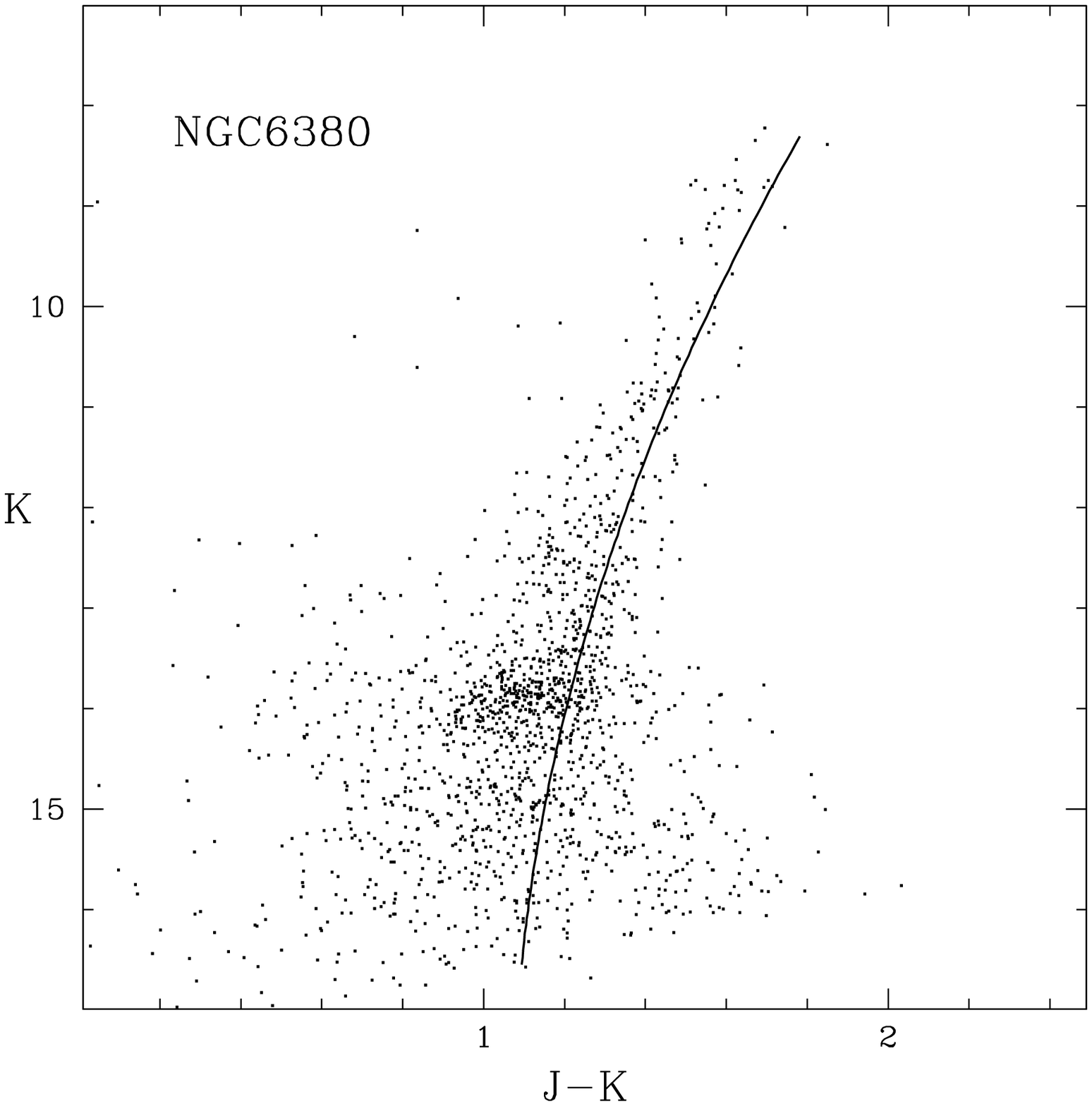}{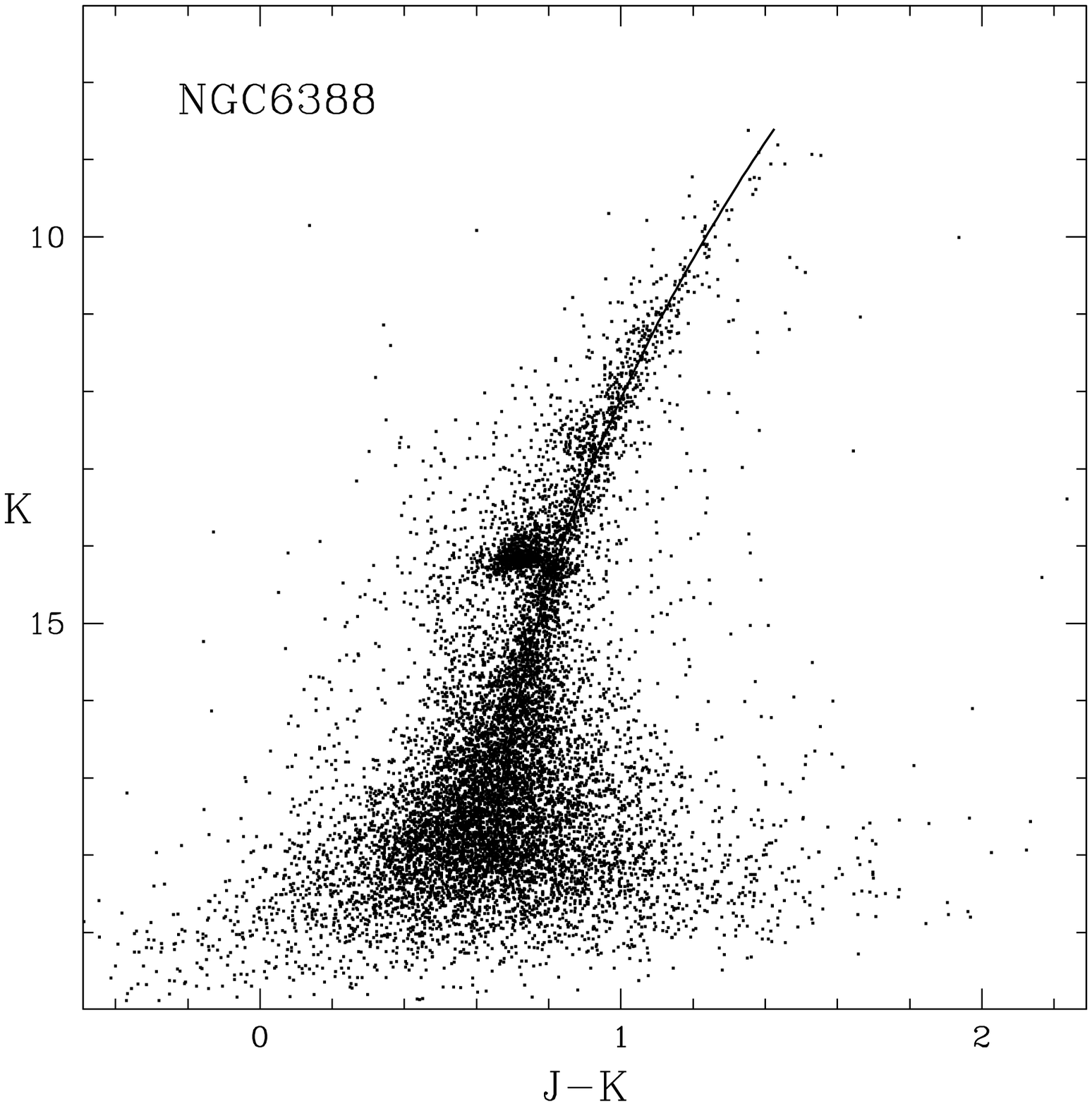}
\plottwo{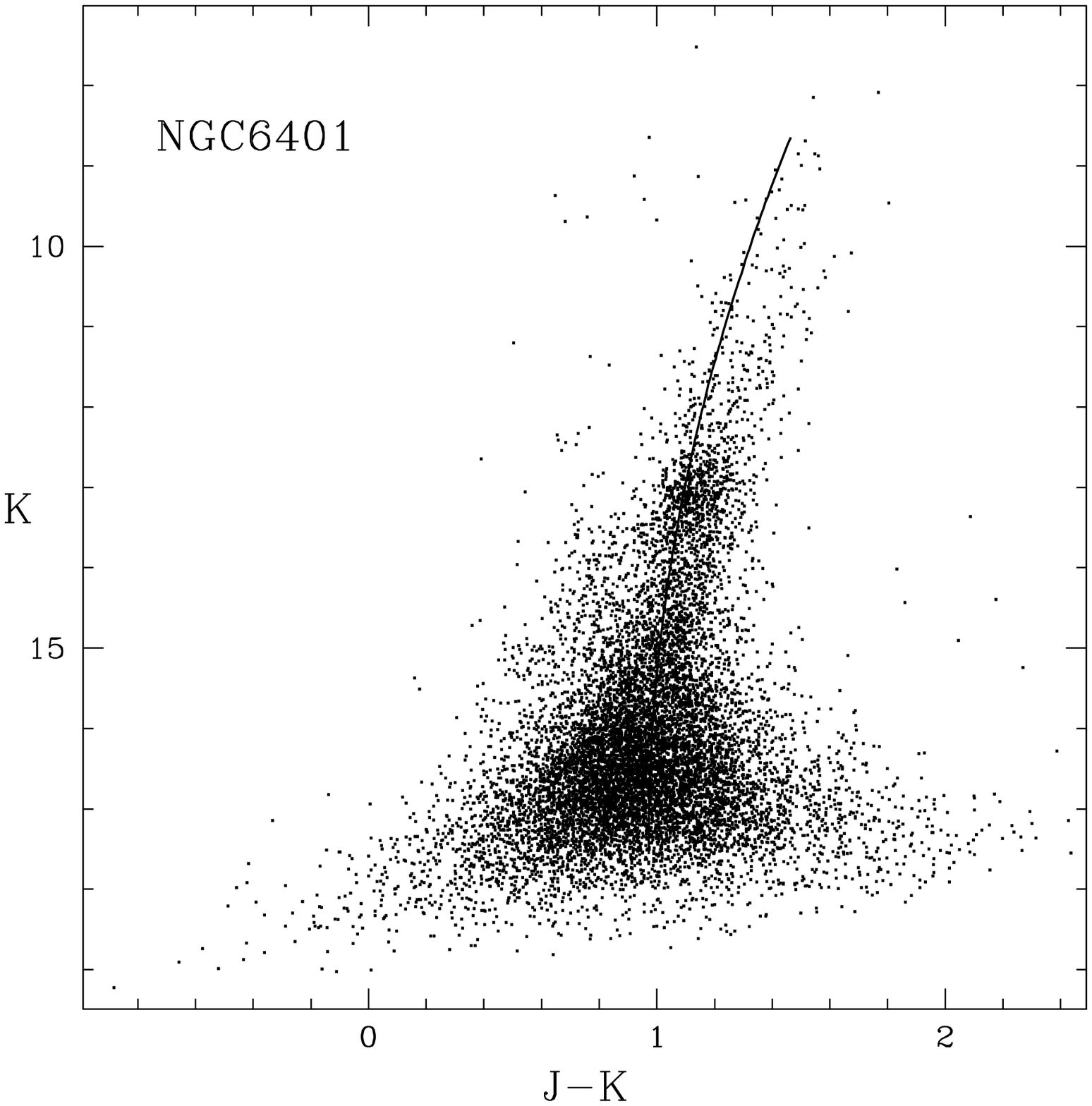}{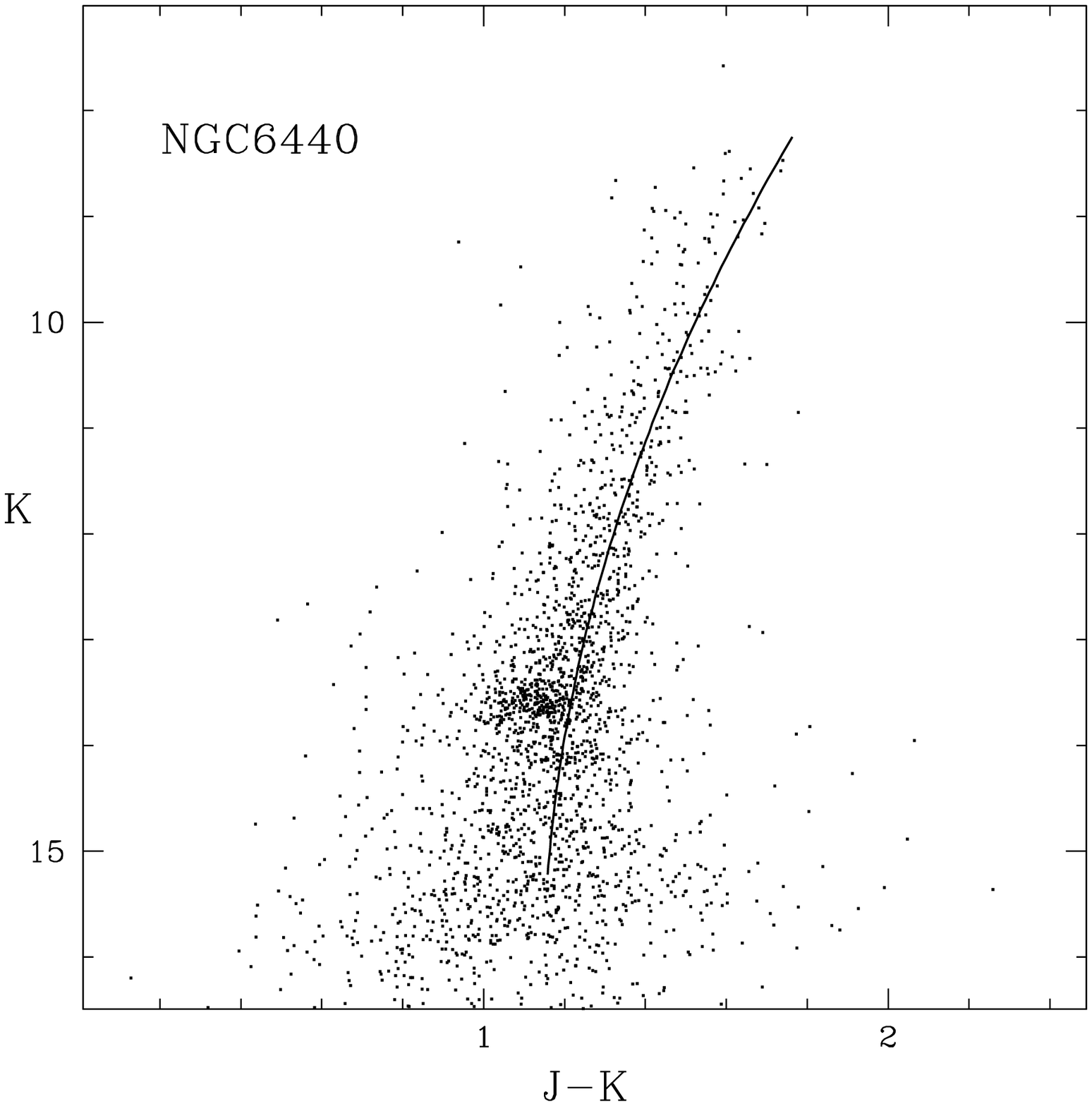}
\caption{continued.}
\end{figure}

\begin{figure}
\setcounter{figure}{1}
\plottwo{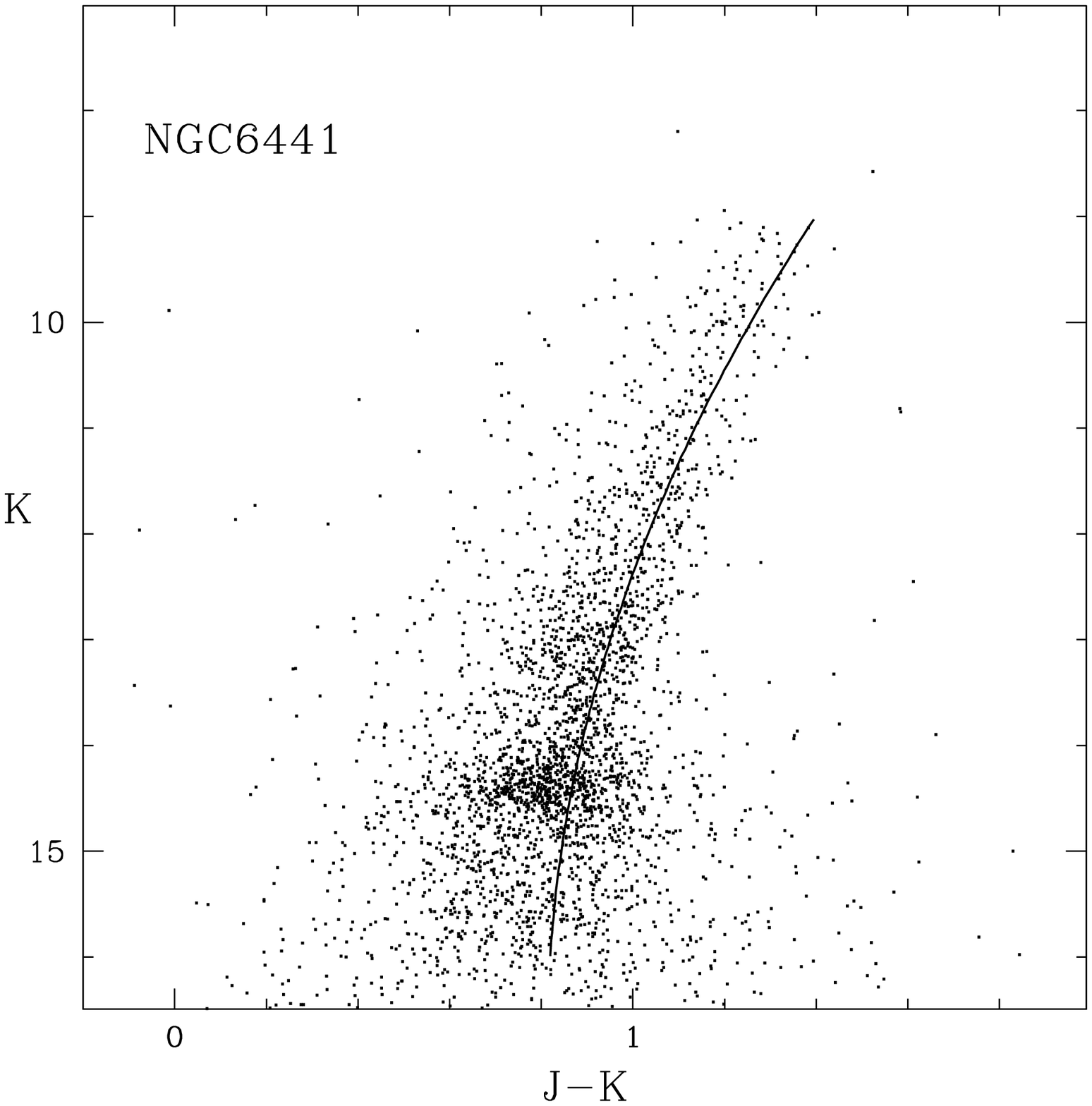}{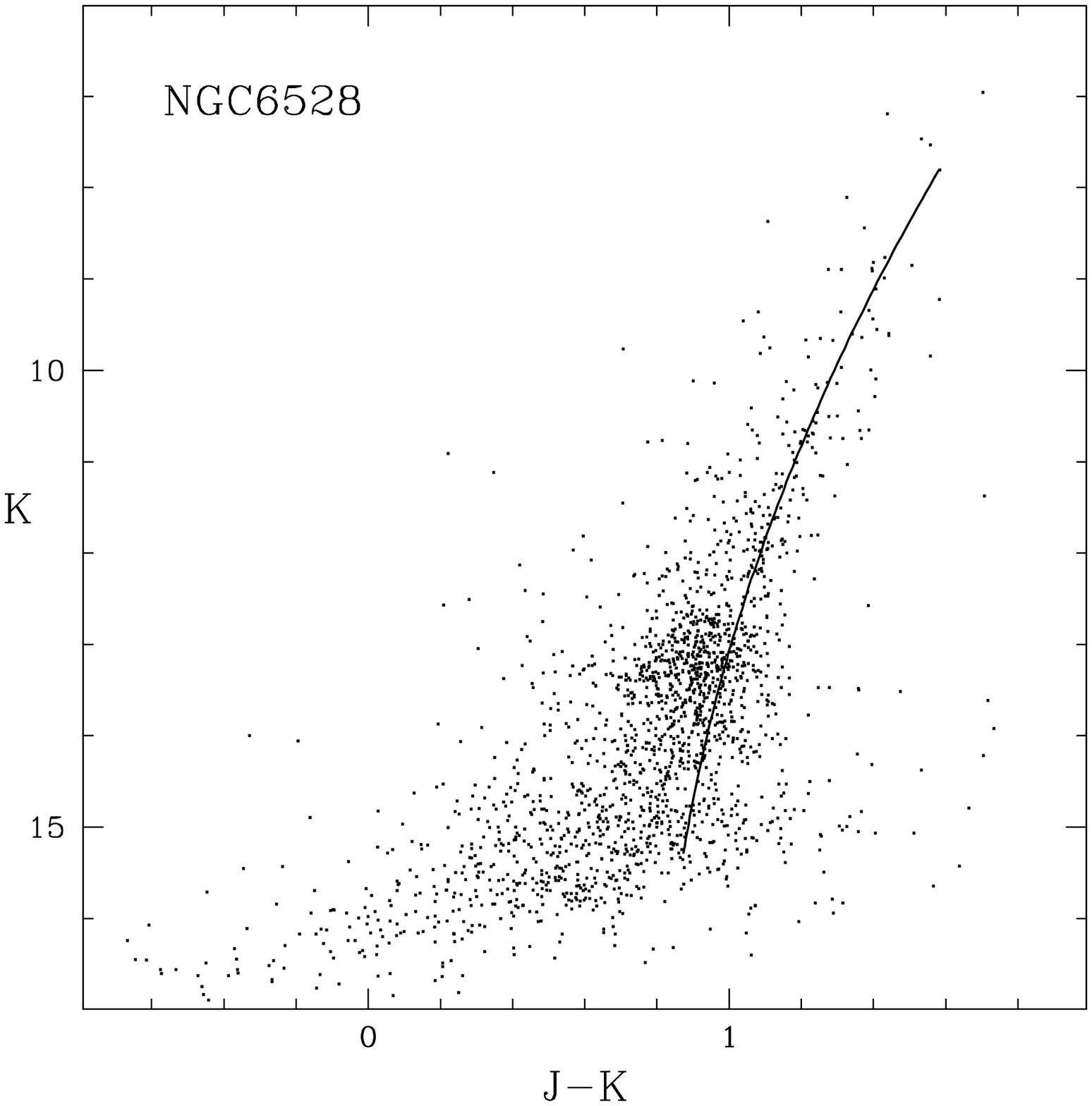}
\plottwo{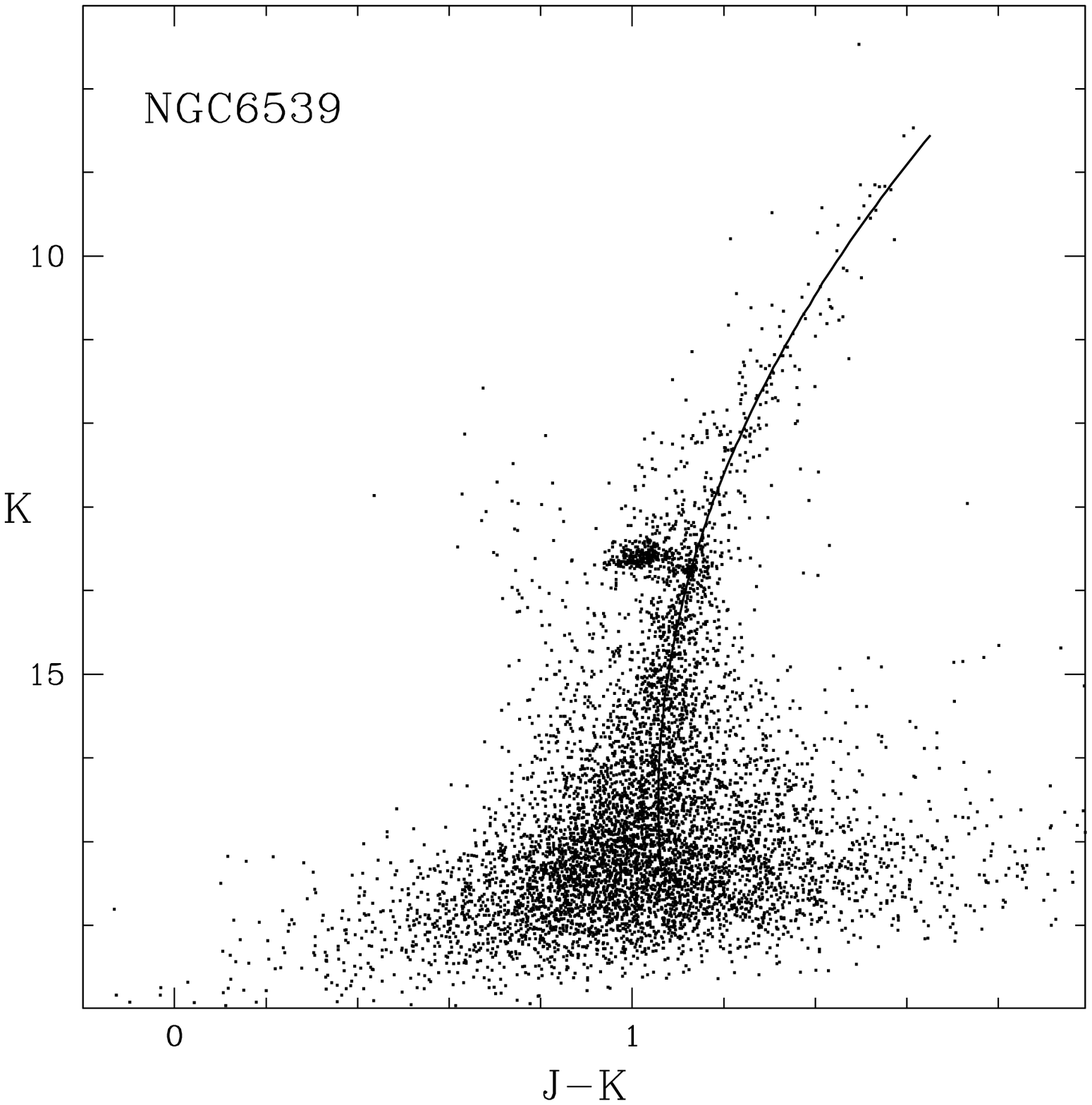}{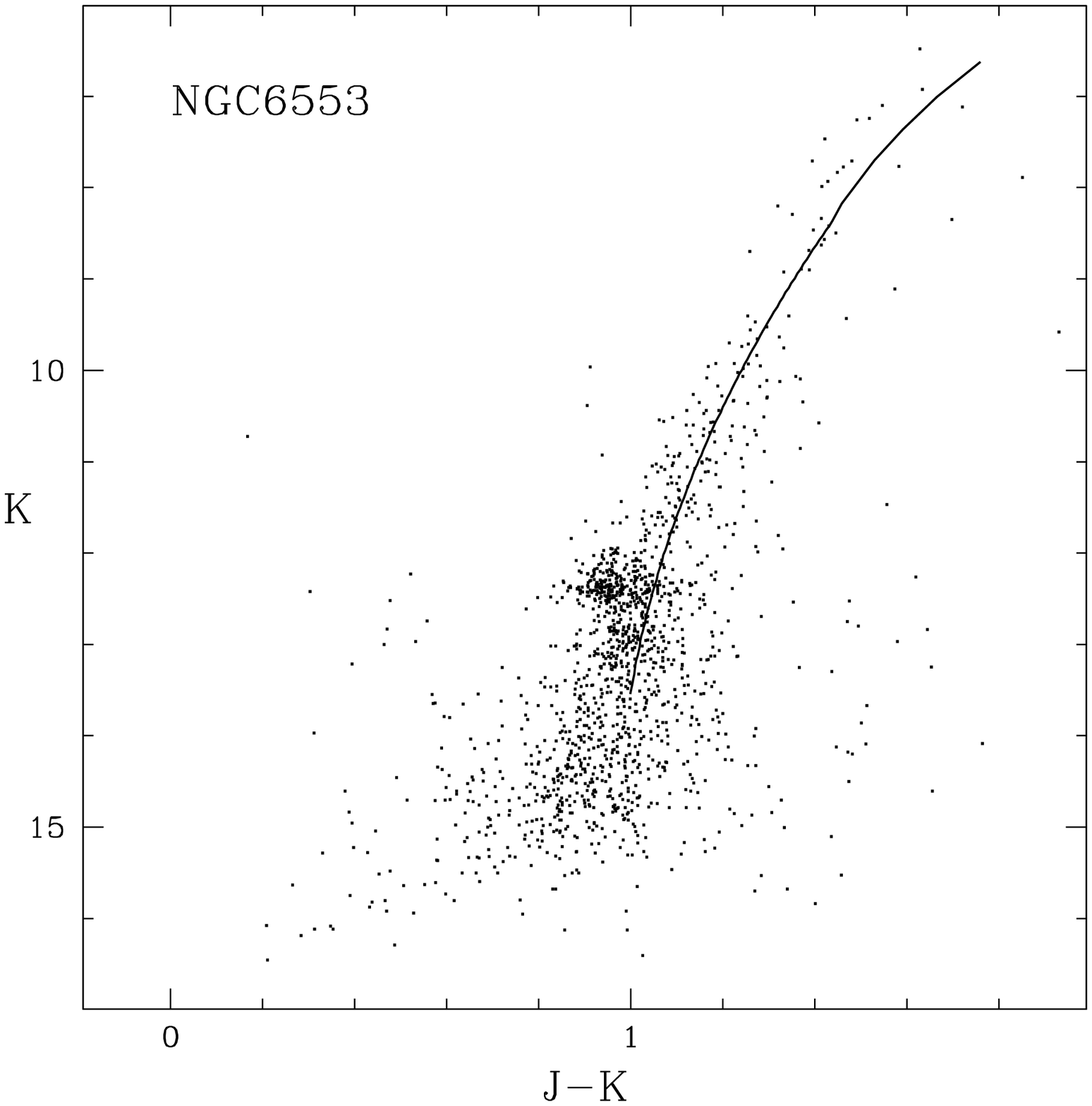}
\plottwo{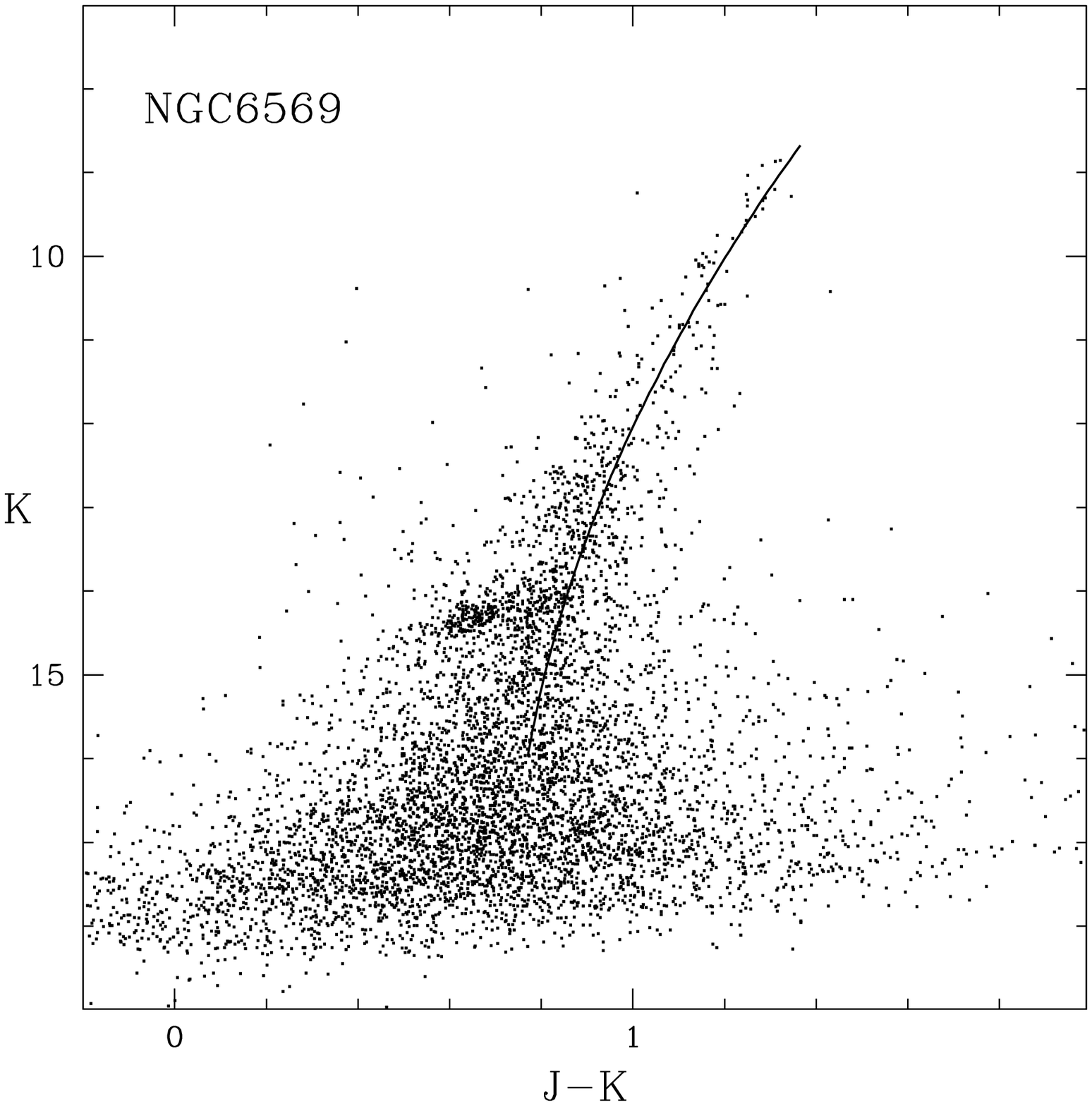}{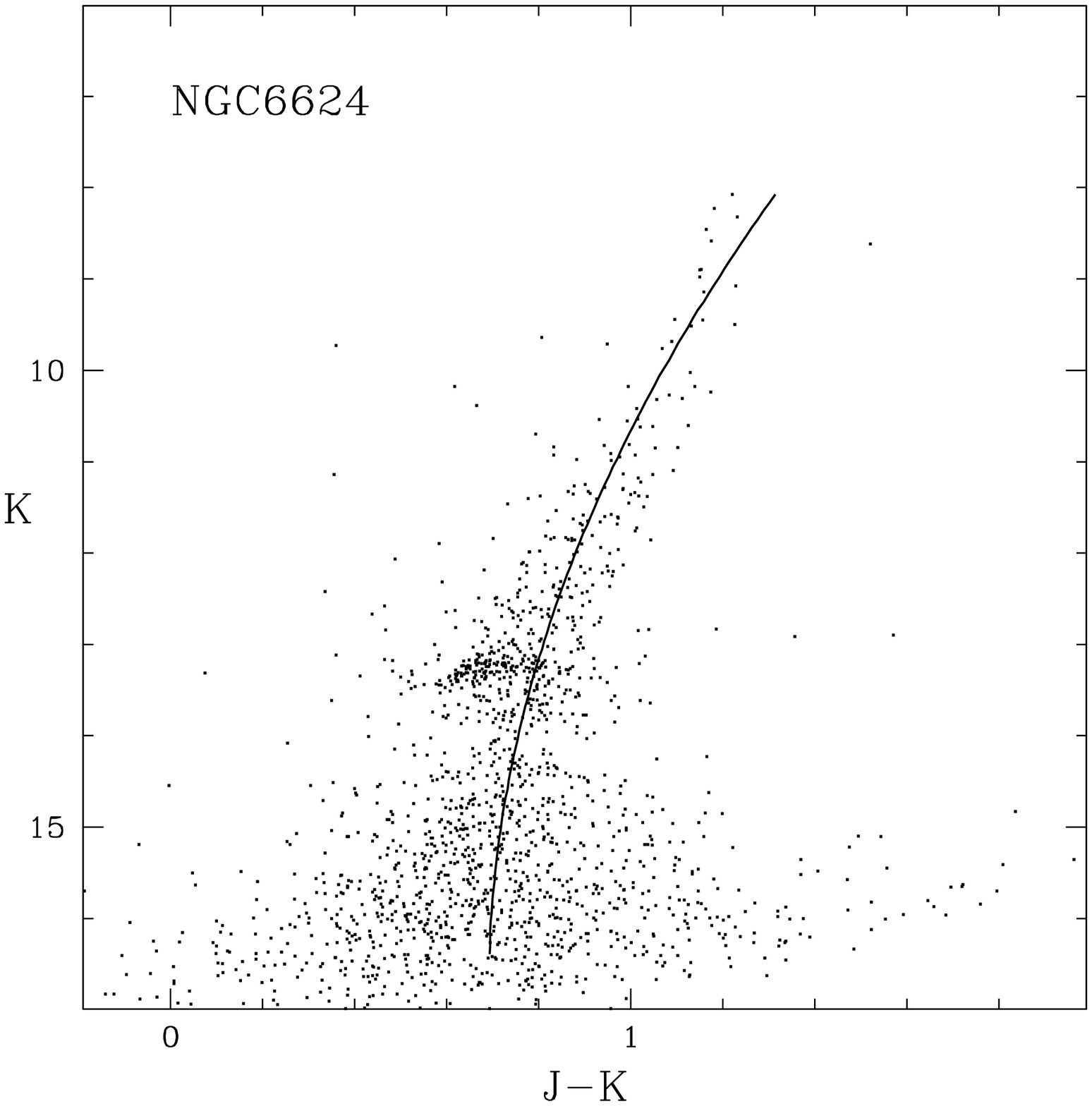}
\caption{continued.}
\end{figure}

\begin{figure}
\setcounter{figure}{1}
\plottwo{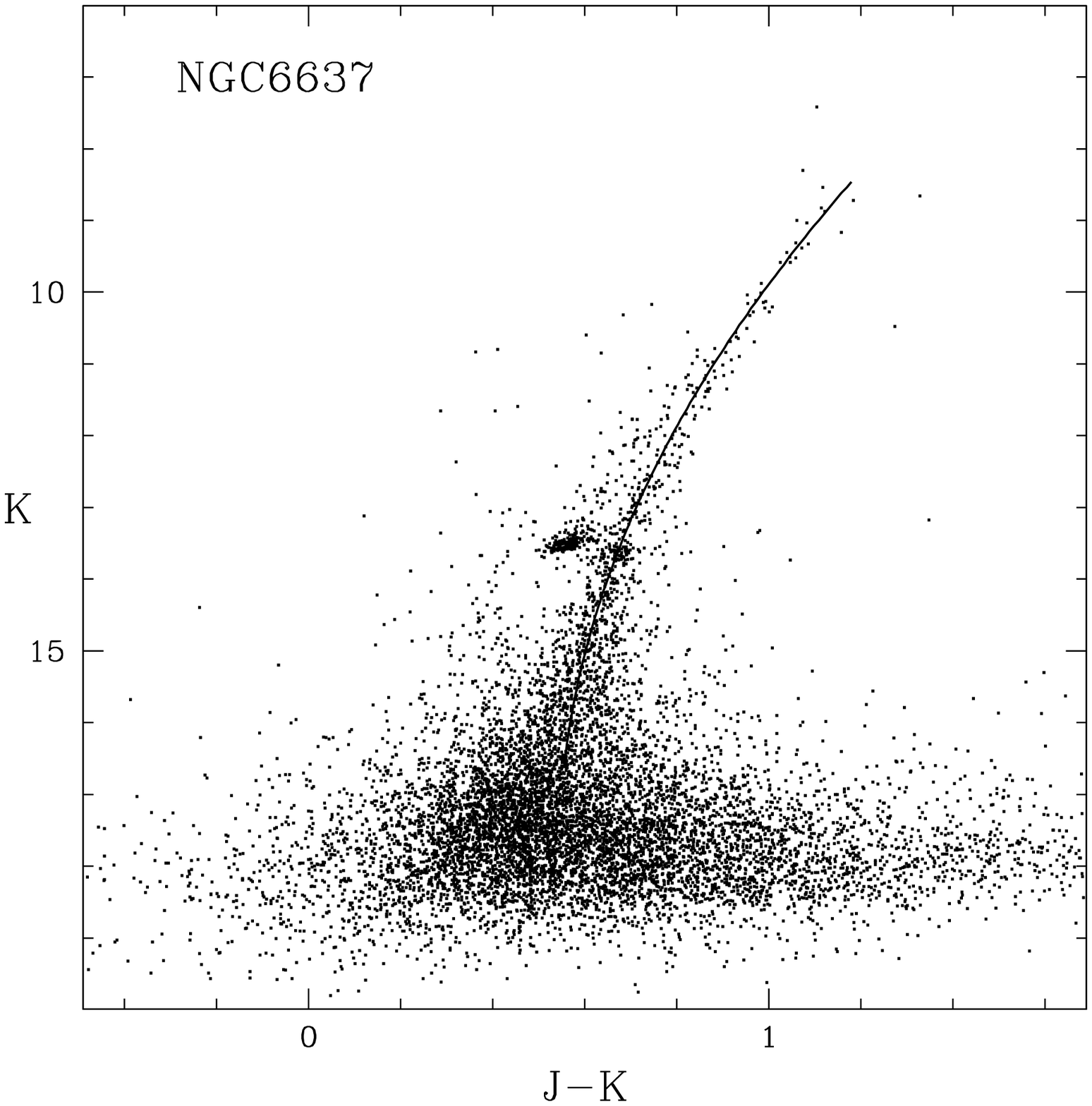}{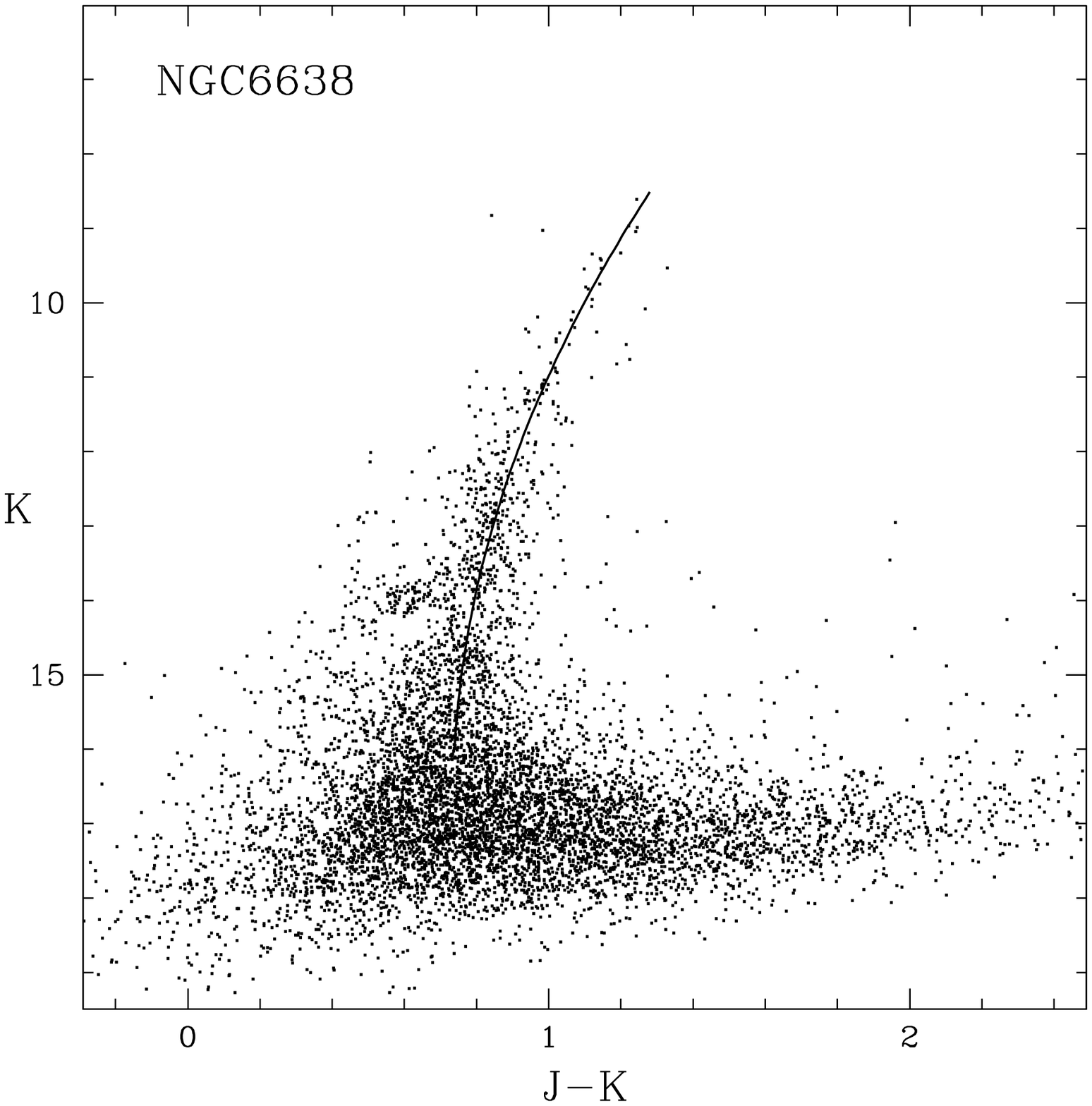}
\plottwo{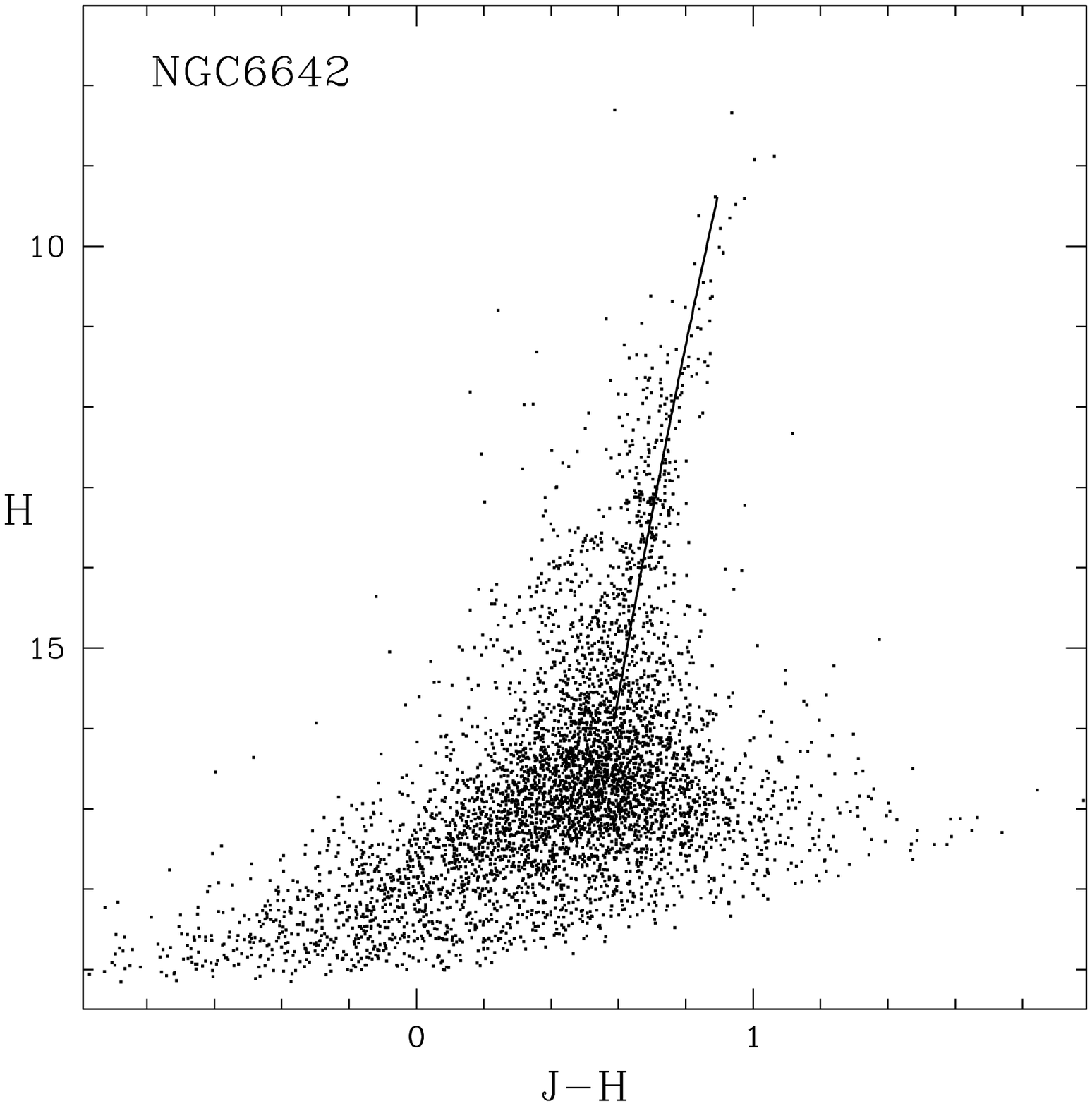}{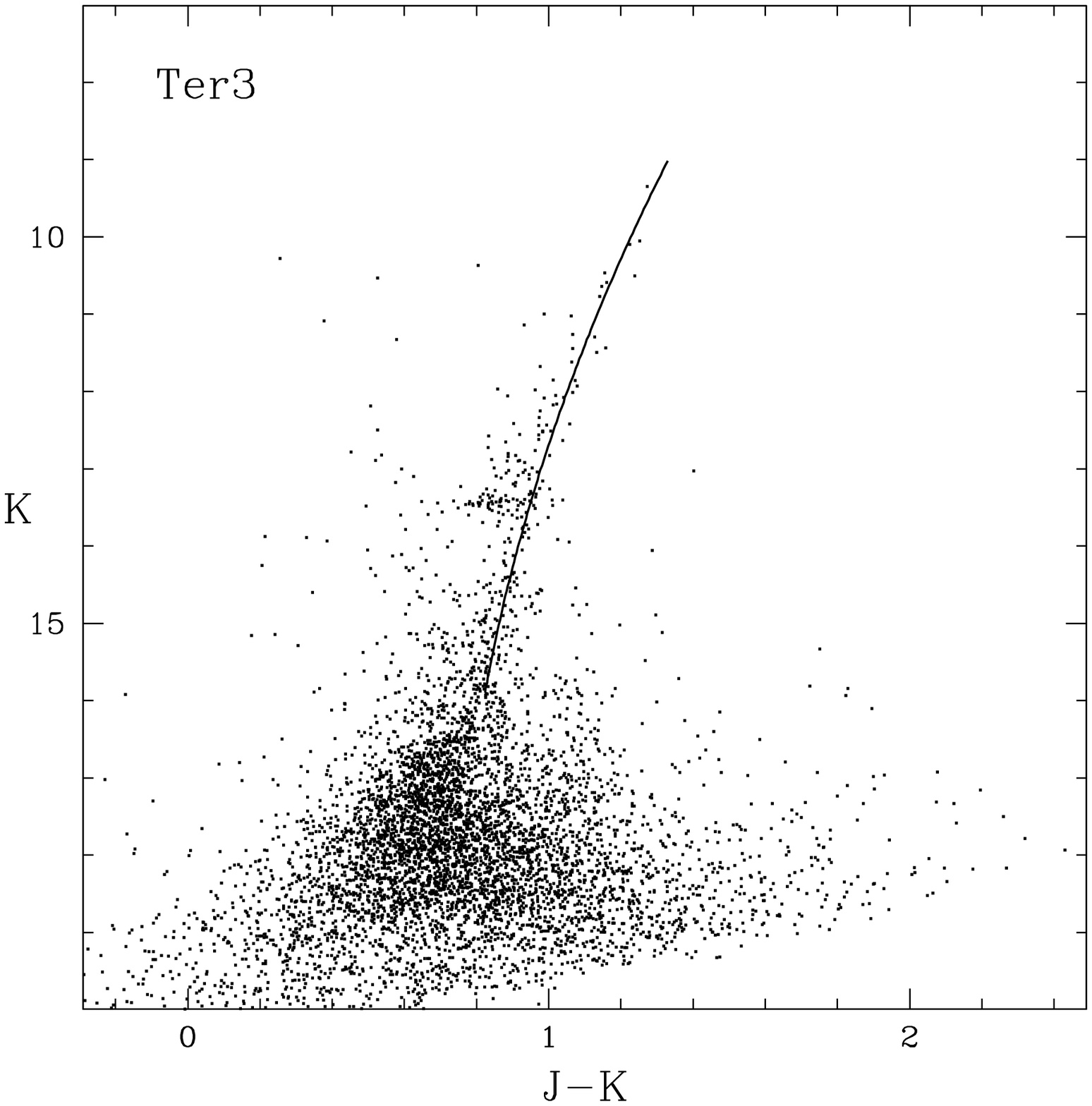}
\plottwo{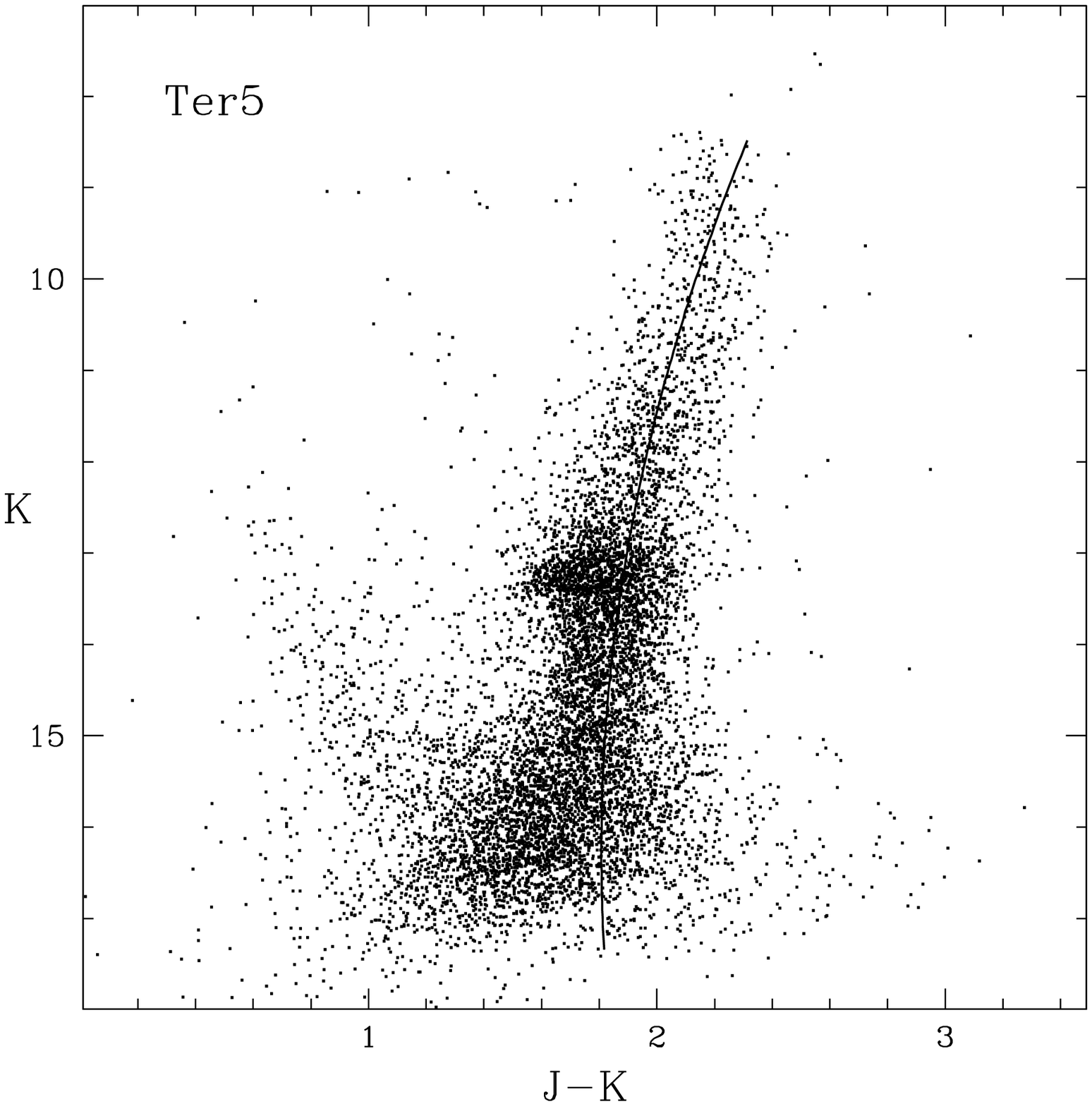}{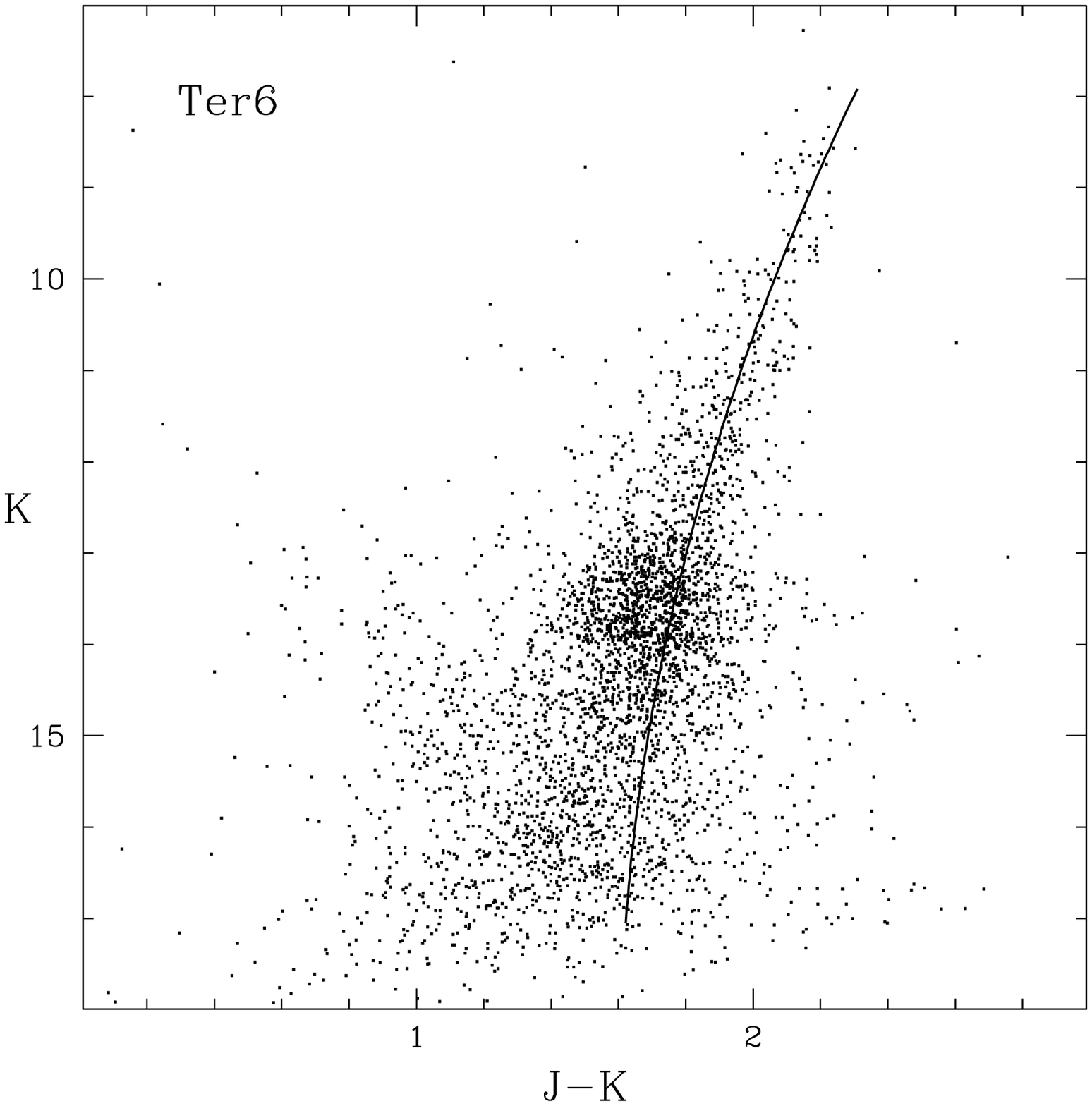}
\caption{continued.}
\end{figure}

\end{document}